\documentclass[journal=jpccck,manuscript=article]{achemso}

\usepackage[utf8]{inputenc}
\usepackage{siunitx}
\usepackage{stfloats}
\usepackage{amsmath}
\usepackage{amsfonts}
\usepackage{amssymb}
\usepackage{graphicx}
\usepackage{natbib}
\usepackage{hyperref}
\usepackage[textsize=tiny]{todonotes}
\usepackage{booktabs}
\usepackage{url}
\usepackage{float}
\usepackage{subfloat}
\usepackage{multirow}
\usepackage{subcaption}
\usepackage{duckuments}
\usepackage[version=3]{mhchem} %
\usepackage{xcolor,colortbl}

\definecolor{Gray}{gray}{0.95}
\definecolor{LightGray}{gray}{0.8}

\newcolumntype{a}{>{\columncolor{Gray}}c}
\newcolumntype{b}{>{\columncolor{white}}c}
\title{Development and Application of a ReaxFF Reactive Force Field for Ni-Doped MoS$_2$}

\author{Karen~Mohammadtabar}
\affiliation{Department of Mechanical Engineering, University of California Merced, 5200 N. Lake Road, Merced, California 95343, United States}
\author{Enrique~Guerrero}
\affiliation{Department of Physics, University of California Merced, 5200 N. Lake Road, Merced, California 95343, United States}

\author{Sergio~Romero Garcia}
\affiliation{Department of Mechanical Engineering, University of California Merced, 5200 N. Lake Road, Merced, California 95343, United States}

\author{Yun Kyung Shin}
\affiliation{Department of Mechanical Engineering, Pennsylvania State University, University Park, Pennsylvania 16802, United States}

\author{Adri C. T. van Duin}
\affiliation{Department of Mechanical Engineering, Pennsylvania State University, University Park, Pennsylvania 16802, United States}

\author{David~A.~Strubbe}
\affiliation{Department of Physics, University of California Merced, 5200 N. Lake Road, Merced, California 95343, United States} 

\author{Ashlie~Martini}
\affiliation{Department of Mechanical Engineering, University of California Merced, 5200 N. Lake Road, Merced, California 95343, United States} 
\email{amartini(at)ucmerced.edu}

\begin{document}

\maketitle

\begin{abstract}
The properties of $\mathrm{MoS_2}$ can be tuned or optimized through doping.
In particular, Ni doping has been shown to improve the performance of $\mathrm{MoS_2}$ for various applications, including catalysis and tribology. 
To enable investigation of Ni-doped $\mathrm{MoS_2}$ with reactive molecular dynamics simulations, we developed a new ReaxFF force field to describe this material.
The force field parameters were optimized to match a large set of density-functional theory (DFT) calculations of 2H-$\mathrm{MoS_2}$ doped with Ni, at four different sites (Mo-substituted, S-substituted, octahedral intercalation, and tetrahedral intercalation), under uniaxial, biaxial, triaxial, and shear strain.
The force field was evaluated by comparing ReaxFF- and DFT-relaxed structural parameters, the tetrahedral/octahedral energy difference in doped 2H, energies of doped 1H and 1T monolayers, and doped 2H structures with vacancies.
We demonstrated the application of the force field with reactive simulations of sputtering deposition and annealing of Ni-doped $\mathrm{MoS_2}$ films. 
Results show that the developed force field can successfully model the phase transition of Ni-doped $\mathrm{MoS_2}$ from amorphous to crystalline.
The newly developed force field can be used in subsequent investigations to study the properties and behavior of Ni-doped $\mathrm{MoS_2}$ using reactive molecular dynamics simulations.
\end{abstract}

\section{Introduction}

Molybdenum disulfide ($\mathrm{MoS_2}$) is a transition metal dichalcogenide with a layered structure where each layer consists of molybdenum atoms sandwiched between sulfur atoms.
Three main applications of the material are in catalysis, opto-electronics, and tribology.
The chemically active edges of $\mathrm{MoS_2}$ nanoparticles have been used to catalyze various reactions, including, hydrogen evolution,~\cite{jaramillo2007MoS2-NP-catal-HER} hydrotreatment of oil,\cite{Jeong2018} and pollutant removal.~\cite{araki2002MoS2-pollutant, bhattacharya2020MoS2-NP-pollutant-degredation}
In 2D form, $\mathrm{MoS_2}$ is widely used as a catalyst for hydrogen evolution reactions~\cite{Deng2015,zong2009,Cuddy2014} and $\mathrm{CO_2}$ reduction.~\cite{Mao2018,kumari20202D-MoS2-Co2Reduction,li2021heterojunction-MoS2-Co2Reduction,chen2020MoS2-Co2degradation}
In opto-electronics, single-layer $\mathrm{MoS_2}$ is a semiconductor with a direct band gap that can be used to construct high efficiency transistors.~\cite{Radisavljevic2011,Yuan2015MoS2Transistor} 
$\mathrm{MoS_2}$ can be synthesized by chemical vapor deposition to form large-area monolayers for use as atomically thin optical and photovoltaic devices.~\cite{Zande2013}
In tribology, the weak van der Waals forces between layers provide low shear resistance, making $\mathrm{MoS_2}$ an effective low-friction solid lubricant or, in nanoparticle form, a liquid lubricant additive.~\cite{Vazirisereshk2019,Dominguez2017MoS2solidLubricant,guo2020MoS2-NP-additive,mousavi2020experimental-MoS2-additive-NP} 

$\mathrm{MoS_2}$ can be doped to enhance its properties.~\cite{Vazirisereshk2019}
Many different elements have been explored as possible $\mathrm{MoS_2}$ dopants, particularly transition metals.~\cite{Vazirisereshk2019,Lauritsen2007,Zhang2017}
Dopants can provide various benefits, including altering the band gap,\cite{Ko2016MoS2electrical} catalytic reactivity,\cite{Xie2016MoS2catal} hardness,\cite{Azhar2020Nidoped} and nanoscale friction.\cite{Acikgoz_2022}
Here, we focus on Ni dopants, which have been shown to improve the performance of $\mathrm{MoS_2}$ for various applications.~\cite{kong2018HER-Ni-dopedMoS2-experimental,rajendhran2018Ni-doped-MoS2-Tribology-enhancement,Jeong2018}
DFT simulations have found four meta-stable sites for a Ni dopant atom in the 2H-$\mathrm{MoS_2}$ crystallographic structure.~\cite{guerrero2020phase}
Dopants can replace an Mo or an S atom in the crystal structure, or they can be intercalated between $\mathrm{MoS_2}$ layers, either between a sulfur site in one layer and a molybdenum site in the other layer (tetrahedral with 4 Ni--S bonds), or between hexagonal holes in both layers (octahedral with 6 Ni--S bonds).~\cite{karkee2021structural}

Studies have investigated the atomistic structure of Ni-doped $\mathrm{MoS_2}$ as well as its electronic and tribological properties.
In doped $\mathrm{MoS_2}$ nano-clusters, the Ni dopant was reported to substitute Mo atoms at edge sites leading to truncation of the cluster morphology relative to un-doped $\mathrm{MoS_2}$.~\cite{Lauritsen2007}
Conversely, Ni doping has also been found to enhance $\mathrm{MoS_2}$ crystal size by increasing the mobility of edge planes during crystallization.~\cite{Kondekar2019}. It has been observed that Ni doping can also transform the 2H-$\mathrm{MoS_2}$ structure to the metallic 1T phase.~\cite{pan2020Ni-2H-->1T,jiang2021Ni-2H-->1T}

Studies have shown that Ni doping increases the number of active sites which, in turn, improves the catalytic performance of $\mathrm{MoS_2}$ in reduction of graphene oxide~\cite{Geng2017}, gas sensing~\cite{Zhang2017}, and hydrogen evolution and production.~\cite{Lopez2016,Wang2016,Zhang2016,kong2018HER-Ni-dopedMoS2-experimental}
Ni also increases the S-vacancy defect density,~\cite{xu2022Ni-doping-S-defect-density} resulting in better catalytic activity for hydrogen evolution reaction.~\cite{dong2020Ni-doping-electronic}
Ni doping changes the electronic properties of $\mathrm{MoS_2}$~\cite{khan2020Ni-doping-electronic, dong2020Ni-doping-electronic, xuan2018Ni-doping-electronic}:
specifically, doping enhances the low electrical conductivity of $\mathrm{MoS_2}$, making this material a promising candidate for electronic applications such as batteries.~\cite{zhang2021Ni-doping-effect,jenisha2022Ni-doping-effect-Electrical} 
In tribology, it has been shown that $\mathrm{MoS_2}$ films co-sputtered with Ni compare favorably to un-doped $\mathrm{MoS_2}$ in terms of friction, wear, and useful life of mechanical parts.~\cite{Stupp1981,Rajendhran2018,Zabinsky1995,Azhar2020Nidoped} 
The improvement in the tribological performance of $\mathrm{MoS_2}$ is particularly notable at low temperatures, which makes Ni-doped $\mathrm{MoS_2}$ ideal as a solid lubricant for space applications where performance at extreme conditions is critical.~\cite{Hamilton2008,Zou2006}

Ni-doped $\mathrm{MoS_2}$ has been studied using \textit{ab initio} density functional theory (DFT) calculations.
Such calculations have shown that the activity of edge sites is doubled~\cite{Wang2015} and that gas adsorption and sensing is enhanced~\cite{Wei2018,MaD2016} by Ni.
Other studies showed that Ni doping improves the catalytic performance of $\mathrm{MoS_2}$ by decreasing the surface sulfur-metal bonding energy~\cite{Raybaud2000}, as well as weakening the S--H bond strength.~\cite{hao2019DFT-Ni-MoS2-catalytic-Performance-HER}
Previous DFT-based studies have provided details about structures, bonding, thermodynamics, vibrational properties, elasticity, and interlayer binding in Ni-doped bulk 2H, bulk 3R, and monolayer  1H-$\mathrm{MoS_2}$ in different phases.~\cite{guerrero2020phase, karkee2021structural} DFT studies have also examined the energies and structural changes in frictional sliding of Ni-doped 2H and bilayer MoS$_2$,~\cite{Guerrero_sliding} and the range of different reconstructed phases accessible by Ni-doping of monolayer 1T-MoS$_2$.~\cite{Karkee_panoply}
However, such calculations are computationally demanding, limiting the time- and size-scales of model systems that can be studied.

An alternative simulation approach is molecular dynamics (MD) based on empirical models, or force fields, that describe the interactions between atoms.
Several force fields have been developed, or optimized, for $\mathrm{MoS_2}$.
First, a Stillinger-Weber force field was developed for $\mathrm{MoS_2}$ and used to calculate mechanical and thermal properties of single layer $\mathrm{MoS_2}$.~\cite{jiang2013SW-forcefield-MoS2,Zhou2017SW-FF}
However, the force field could not capture the behavior of $\mathrm{MoS_2}$ at states far from equilibrium,~\cite{Ostadhossein2017} and did not include parameters for interlayer interactions. 
A custom, interpretable force field compatible with various non-reactive potential formalisms was developed for 2H-$\mathrm{MoS_2}$~\cite{Liu2020interpretable} as well as for Cr-doped 2H-$\mathrm{MoS_2}$.~\cite{Xing2022Crdoped}
Despite accurate representation of energetic, mechanical, and surface properties, these potentials do not model the formation and breaking of chemical bonds.
A many-body Mo/S potential based on the Reactive Empirical Bond Order (REBO) and Tersoff potentials was developed for $\mathrm{MoS_2}$.~\cite{liang2009REBO-forcefield-MoS2}
The force field was able to reproduce expected lattice constants as well as mechanical properties of $\mathrm{MoS_2}$, but it was unable to accurately model surface energy.
Several force fields within the ReaxFF formalism, that captures the formation and breaking of chemical bonds, have been parameterized for $\mathrm{MoS_2}$ as well.~\cite{Ostadhossein2017,Hong2017,chen2020formation-MoS2-from-elemental-Mo-S, ponomarev2022ReaxFF-MoS2-new}
These force fields have been used in simulations of crystallization,~\cite{Ostadhossein2017,Hong2017,chen2020formation-MoS2-from-elemental-Mo-S,Chen2020DomainO, ponomarev2022ReaxFF-MoS2-new} active edge sites,~\cite{Hu2020} creation of vacancies,~\cite{Ostadhossein2017,Yilmaz2018,noori2021Mechanical-MoS2-ReaxFF} distribution and dynamics of defects,~\cite{Patra2018} mechanical properties of $\mathrm{MoS_2}$ monolayer heterostructures,~\cite{zahedi2022mechanical-MoS2-ReaxFF,mejia2021MoS2-NanoTubes-ReaxFF,mortazavi2016MoS2-mechanical-properties} and tribological behavior of multi-layer $\mathrm{MoS_2}$.~\cite{Shi2019}
However, to investigate Ni-doped $\mathrm{MoS_2}$, force field parameters that include the interactions between Ni and $\mathrm{MoS_2}$ are needed, which is a challenge as new interactions are introduced and the dopant's effect on the otherwise weak interlayer interactions must be described. 

In this study, two new ReaxFF force fields were developed for Ni-doped $\mathrm{MoS_2}$.
The force field parameters were optimized by comparing ReaxFF energies to those obtained from a large set of DFT calculations of the equation of state (energy vs. strain) of Ni-doped 2H-$\mathrm{MoS_2}$ under uniaxial, biaxial, triaxial, and in-plane shear strain.
DFT calculations were performed with Ni dopants at each of four different sites: Mo-substituted, S-substituted, octahedral, and tetrahedral intercalation.
The resulting ReaxFF force field was evaluated based on relaxed bond lengths and structural parameters in 2H, as well as calculations of structures not in the training set such as doped 1H and 1T monolayers and doped 2H with vacancies as well as sliding. Finally, we applied the new force field to model sputter deposition and annealing of Ni-doped MoS$_2$, pointing the way to future applications.

\section{Methods}

\subsection{DFT Calculations}
As in our previous work on thermodynamics and vibrational properties of Ni-doped $\mathrm{MoS_2}$\cite{guerrero2020phase}, the plane-wave density functional theory (DFT) code Quantum ESPRESSO\cite{Giannozzi2017} was used for quantum calculations. 
The Perdew-Burke-Ernzerhof\cite{Perdew1996} (PBE) generalized gradient approximation was used with Grimme-D2\cite{Grimme2006} van der Waals correction, and the electron-ion interaction was described with optimized norm-conserving Vanderbilt pseudopotentials\cite{Hamann2013} parametrized by Schlipf and Gygi.\cite{Schlipf2015} 
All DFT computations used a kinetic energy cutoff of 60~Ry. 
PBE + Grimme-D2 has been shown to accurately describe the lattice parameters, elastic constants, and phonon frequencies of $\mathrm{MoS_2}$.\cite{guerrero2020phase} Ni-doped 2H MoS$_2$ has been previously computed not to be magnetic,\cite{karkee2021structural} so we used non-spin-polarized DFT.

\begin{figure}[H]
\centering
\includegraphics[width=.95\textwidth]{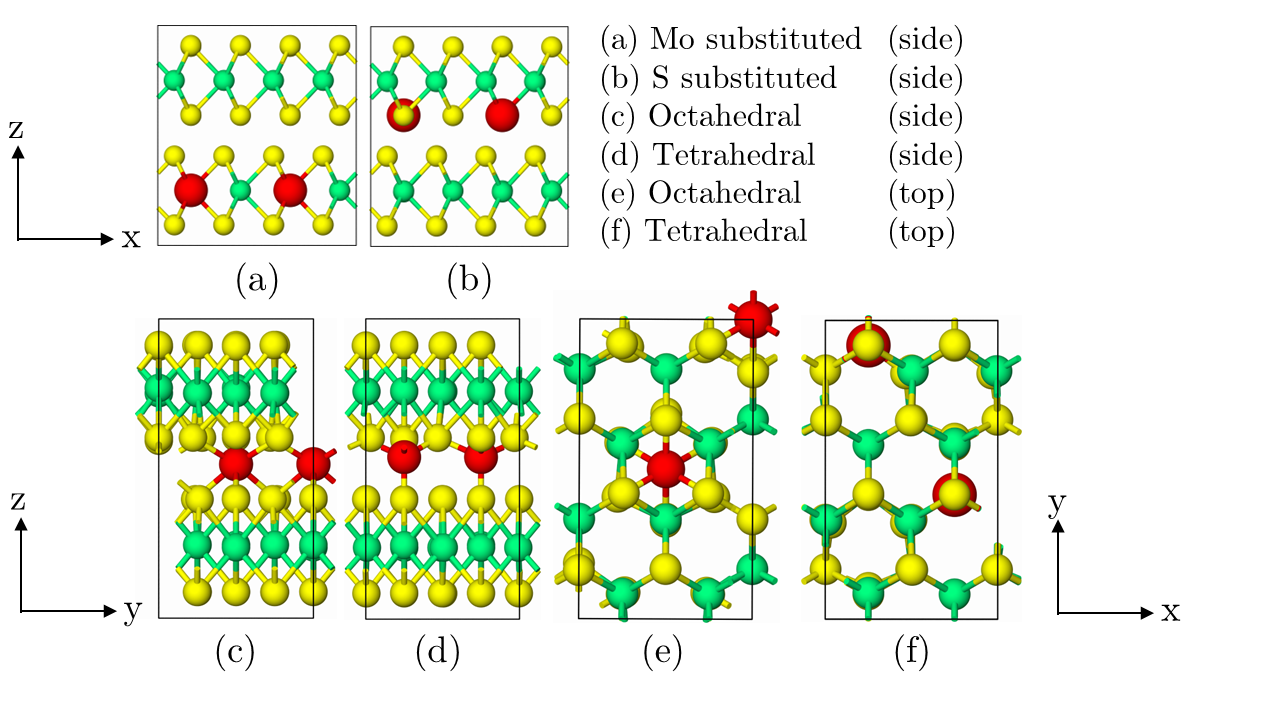}
\caption{Side views of the training set structures, illustrating the four possible locations of the Ni dopant in $\mathrm{MoS_2}$: (a) Mo-substituted, (b) S-substituted, (c) octahedral intercalation, and (d) tetrahedral intercalation. (e) and (f) show top views of the octahedral and tetrahedral intercalations, respectively. Sphere colors correspond to S (yellow), Mo (green), and Ni (red).}
\label{fig:model}
\end{figure}

The training set was composed of  2H-$\mathrm{MoS_2}$ in $2\times2\times1$ supercells, where the third direction is perpendicular to the basal plane of the layers. While there are effects of doping concentration, as studied in detail in \citet{guerrero2020phase}, this size of supercell is sufficient to capture the local interactions that the force field must describe.
Pristine 2H-bulk structures have six atoms per unit cell; a half-shifted Monkhorst-Pack $k$-grid of $4\times4\times4$ was used. In the absence of any available structures from X-ray diffraction, our starting configurations are the relaxed pristine structures with Ni atoms substituted, or intercalated in the high-symmetry tetrahedral or octahedral locations. The structures were relaxed toward zero stress using Quantum ESPRESSO's standard BFGS algorithm. The relaxation ended when forces and stresses were below thresholds of 10$^{-4}$ Ry/Bohr and 0.1 kbar, respectively. The relaxed lattice parameters of the hexagonal primitive cell are $a = b$ = 3.19 \AA{}, $c$ = 12.40 \AA{}, $\alpha = \beta = 90^\circ$, and $\gamma = 120^\circ{}$.
The dopant sites were chosen because they are stable or meta-stable \cite{guerrero2020phase}; other sites such as intralayer interstitial or S--S bridge intercalation would relax to other structures. 
The stable doped structures (Mo-substituted, S-substituted, and intercalation at the tetrahedral and octahedral sites, shown in Figure~\ref{fig:model}) were taken from \citet{guerrero2020phase}, constructed with one Ni atom in each $2\times2\times1$ supercell, i.e., up to 12.5\% doping (which may be also considered alloying).
For validation, defect calculations in $3\times3\times1$ supercells used a $3\times3\times4$ $k$-grid.
The training set systems are shown in Figure~\ref{fig:model}. Due to lower computational efficiency for systems with acute cell angles in the stand-alone ReaxFF code and LAMMPS, the results were used to construct nearly orthorhombic ($\alpha = \beta = \gamma =\ 90^\circ$) cells with twice the number of atoms and twice the energy, which does not change any intensive properties. All DFT results compared to ReaxFF are in terms of this doubled cell, for clarity.

Strained 2H-bulk structures were studied under six strain conditions: uniaxial $x$-strain, uniaxial $z$-strain, biaxial $xy$-strain, triaxial strain, and $xy$-shear. 
Note that due to exact symmetries in the pristine structure, and approximate symmetries in the doped structures,\cite{guerrero2020phase} uniaxial $y$-strain would not provide further distinct information.
For each strain direction, seven points were sampled with strains ranging from -15\%{} to 15\%{} in intervals of 5\%, consistent with strain ranges in the initial $\mathrm{MoS_2}$ parameterization.\cite{Ostadhossein2017}
Similarly, shear calculations were performed for seven points with shearing angles (the angle between orthogonal $a$ and $b$ lattice vectors between $\sim 72^\circ$ and $108^\circ$.
In each case, the atomic forces were relaxed to $10^{-4}$~Ry/Bohr with fixed lattice vectors.

\subsection{ReaxFF Force Field and Parameterization}

ReaxFF is a reactive empirical force field based on bond order and atomic distances that originally was developed for hydrocarbons.~\cite{vanDuin2001}
Over the years, many different parameter sets have been developed for various chemical systems. 
ReaxFF accounts for the contributions of various partial energy terms that
enable ReaxFF to accurately model covalent and ionic bonds as well as non-bonded interactions. 
The total energy in the force field is the sum of the bond energy, over-coordination and under-coordination energy corrections, angle strain, torsion energy, torsion conjugation, van der Waals, and Coulomb energies. 
A detailed explanation of all terms can be found in the original ReaxFF article.~\cite{vanDuin2001}

We started from two different parameter sets that were previously developed for S/Mo interactions, one reported in 2017~\cite{Ostadhossein2017} and the other reported in 2022.~\cite{ponomarev2022ReaxFF-MoS2-new}
The 2017 potential was developed specifically for single-layer $\mathrm{MoS_2}$, with a focus on its mechanical response with and without vacancies, and included parameters to model interactions between $\mathrm{MoS_2}$ and oxygen.
Then, the 2022 potential was developed by modifying the Mo/S parameters in the 2017 potential to better capture crystallization of $\mathrm{MoS_2}$ in bilayer and bulk form.
We introduced Ni parameters for both the 2017 and the 2022 force fields.
The Ni metal parameter set employed in this work is the same as that of Ni/Mo force field\cite{vasenkov2012Ni-Mo}. The Ni metal force field, including Ni atom parameters (1-body parameters) and Ni-Ni interactions, was originally developed for Ni-catalyzed hydrocarbon chemistry\cite{Mueller2010}. The Ni parameter set was trained against DFT data for the 2017 potential on the equations of states of various crystal phases (fcc, bcc, diamond, A15 and sc), their cohesive energies, and the surface energies for Ni(111) and Ni(100) surfaces.
Since the newer study~\cite{ponomarev2022ReaxFF-MoS2-new} used bulk $\mathrm{MoS_2}$ DFT training data and was demonstrated for crystallization of $\mathrm{MoS_2}$, which is closer to our goal of simulating deposition, we report the results for the new potential based on the 2022 parameters in the main text.
However, comparisons of the training results with the DFT data for the potential based on the 2017 force field (including Ni/S and Mo/Ni training results in Figs. S1-4 and Table S1) as well as both potential files are available as Supporting Information.

Our force field was trained against the DFT data by optimizing the parameters specifically for Mo-S-Ni, S-Mo-Ni, and S-Ni-Mo valence angles. Other parameters, such as those for Ni-S and Mo-Ni bonds, remained fixed to literature values.
The bond angle parameters were the equilibrium angle, first and second force constants, undercoordination parameter, and energy/bond order.
The process of parameterization included calculating the potential energy of each structure ($E^{\rm ReaxFF}$) which was then compared to the energy obtained from DFT for the same structure ($E^{\rm DFT}$).
A weighted error was calculated as:

\begin{equation}
    {\rm Error}=\sum_i \left(\frac{E_i^{\rm ReaxFF}-E_i^{\rm DFT}} {w_i} \right)^2
\label{eq:error}
\end{equation}
\noindent where $w_i$ is the weight associated with each data point on the energy plots. 
The weights were chosen to prioritize minimizing the difference between the DFT and the ReaxFF energies for near-equilibrium structures.
The parameters were optimized by the single-parameter search optimization technique~\cite{van1994delft} in the stand-alone ReaxFF package.
The energy difference between each strained and equilibrated structure as obtained from ReaxFF and DFT was plotted as a function of strain for each strain direction.
The same was done for sheared structures at each shearing angle.
The parameterization process was repeated until the shapes of the energy plots were as similar as possible between ReaxFF and DFT.
This procedure has been used previously to optimize ReaxFF parameters for various chemical systems.~\cite{Ostadhossein2017, Mueller2010, hahn2018development, nayir2019development, Khajeh2019, ponomarev2022ReaxFF-MoS2-new} 
The accuracy of the developed force field was evaluated and it was then used for energy minimization and dynamics simulations with the Large-scale Atomic/Molecular Massively Parallel Simulator (LAMMPS) code.~\cite{LAMMPS}

\section{Results and Discussion}

\subsection{Force Field Parameterization}
\begin{figure}[H]
\centering
\begin{subfigure}[b]{.49\linewidth}
{\includegraphics[width=.95\textwidth]{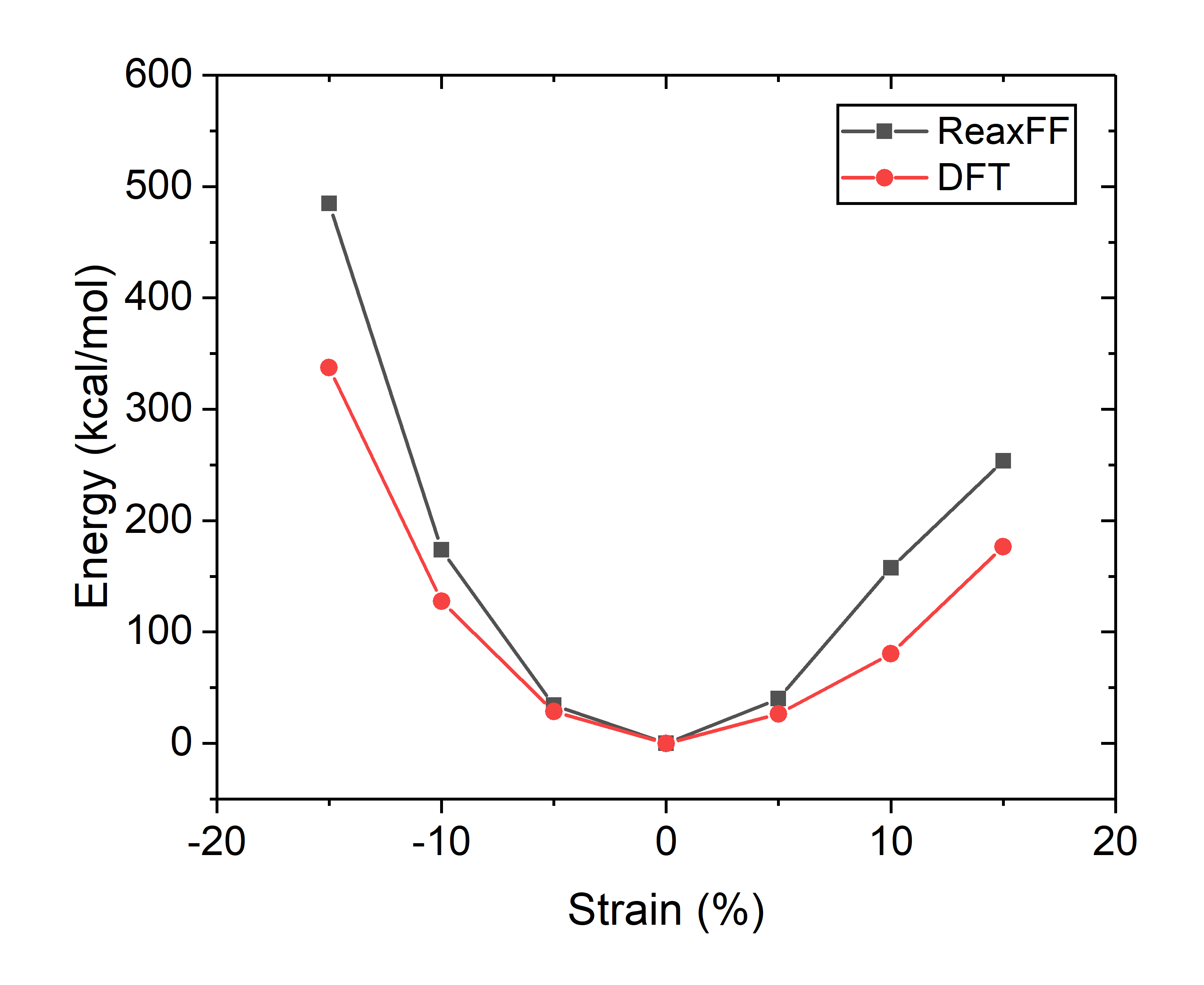}}
\caption{Mo-substituted}
\end{subfigure}
\begin{subfigure}[b]{.49\linewidth}
{\includegraphics[width=.97\textwidth]{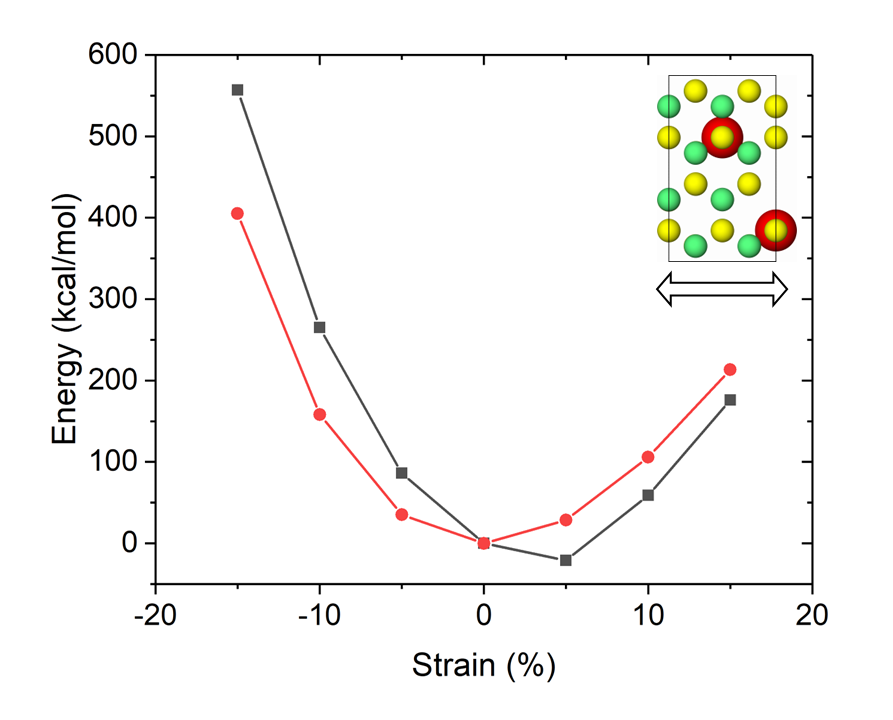}}
\caption{S-substituted}
\end{subfigure}
\begin{subfigure}[b]{.49\linewidth}
{\includegraphics[width=.95\textwidth]{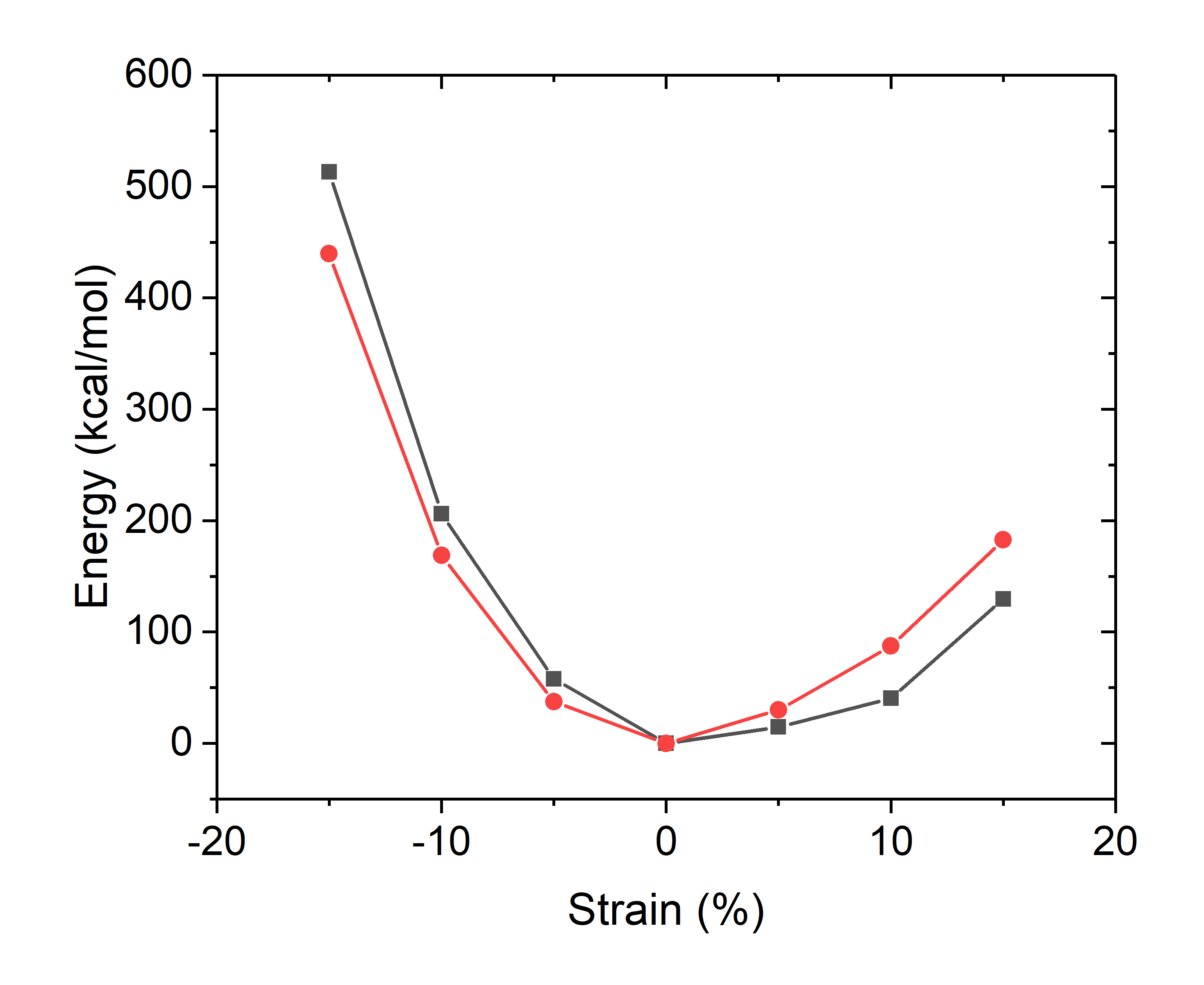}}
\caption{Octahedral}
\end{subfigure}
\begin{subfigure}[b]{.49\linewidth}
{\includegraphics[width=.95\textwidth]{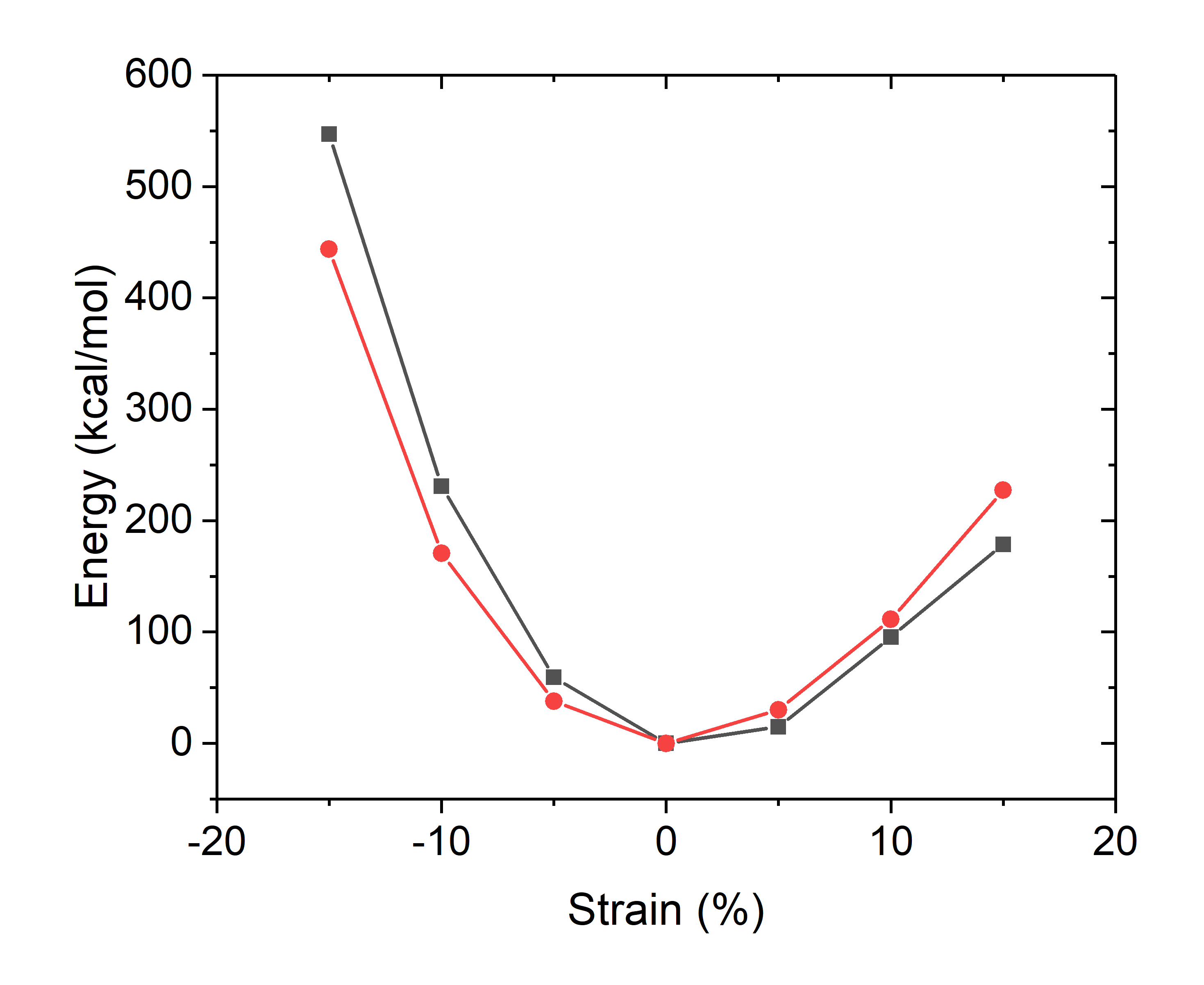}}
\caption{Tetrahedral}
\end{subfigure}
\caption{Equations of state calculated from DFT (red) and ReaxFF (black) for the Ni-doped $\mathrm{MoS_2}$ structures strained uniaxially in the $x$-direction. The inset in (b) shows a top view of the S-substituted structure with an arrow indicating the strain direction. }
\label{fig:EoS-uniX}
\end{figure}

\begin{figure}
\begin{subfigure}[b]{.49\linewidth}
{\includegraphics[width=.95\textwidth]{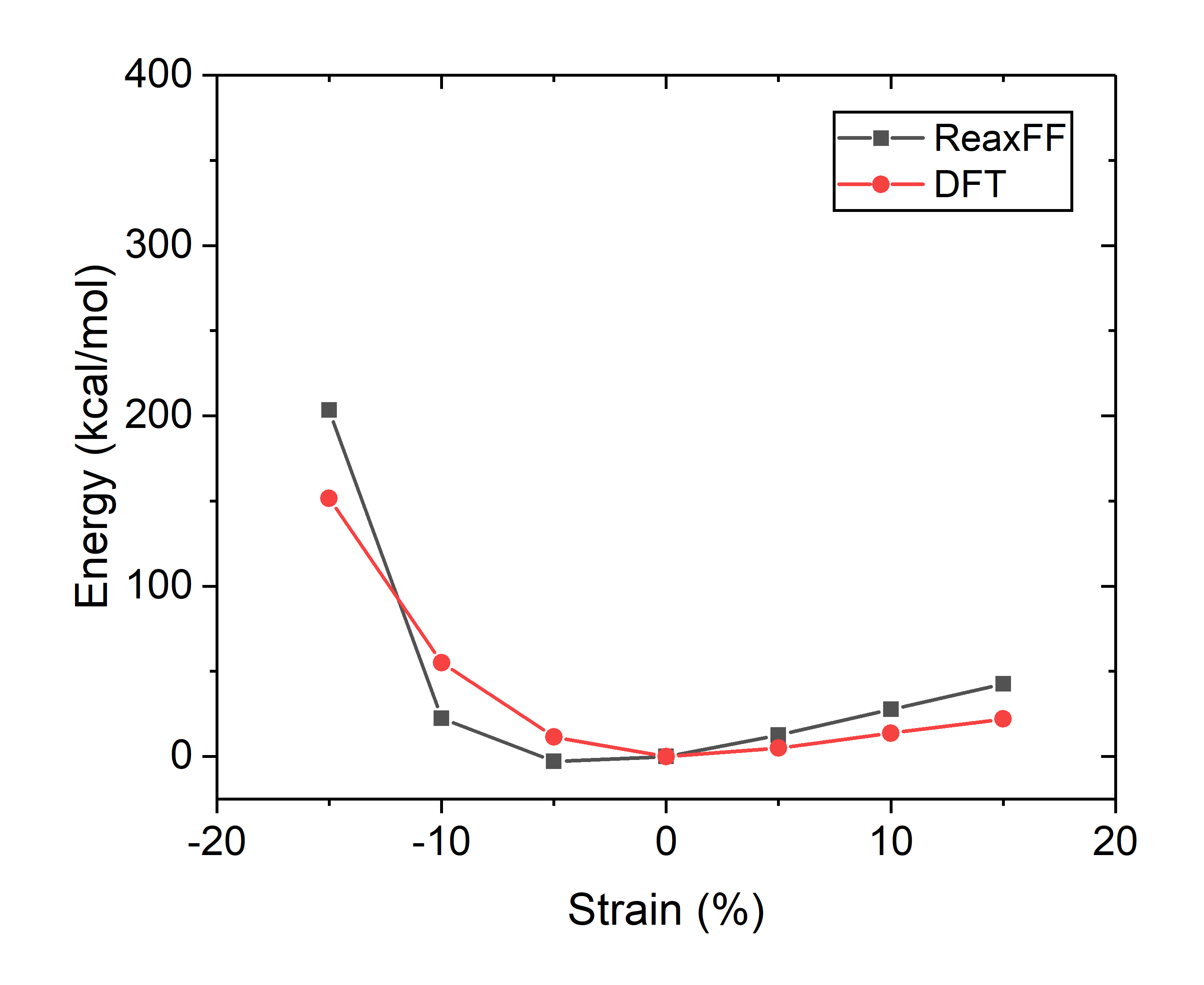}}
\caption{Mo-substituted}
\end{subfigure}
\begin{subfigure}[b]{.49\linewidth}
{\includegraphics[width=.98\textwidth]{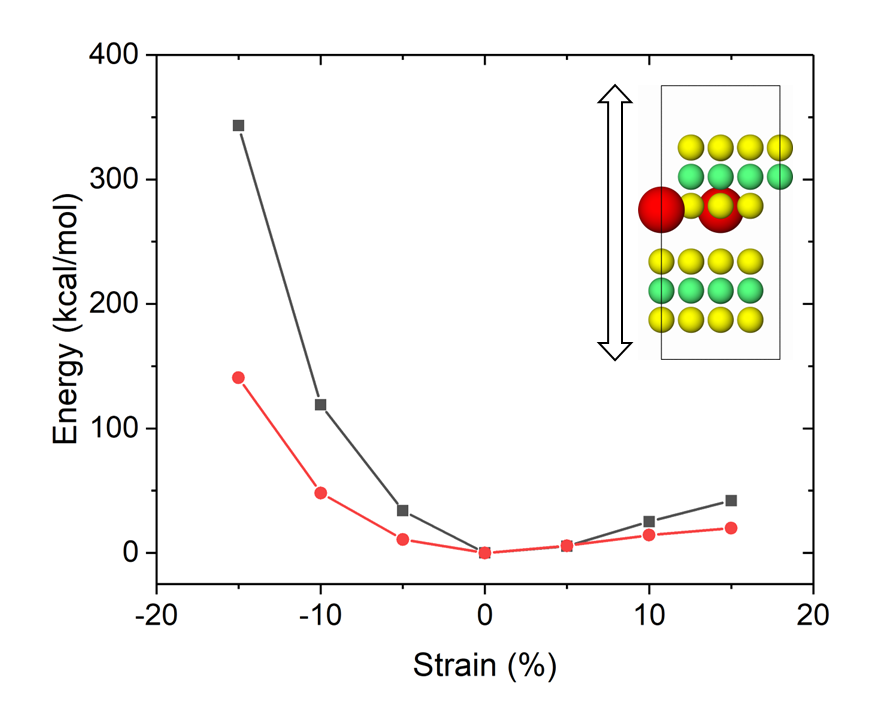}}
\caption{S-substituted}
\end{subfigure}
\begin{subfigure}[b]{.49\linewidth}
{\includegraphics[width=.95\textwidth]{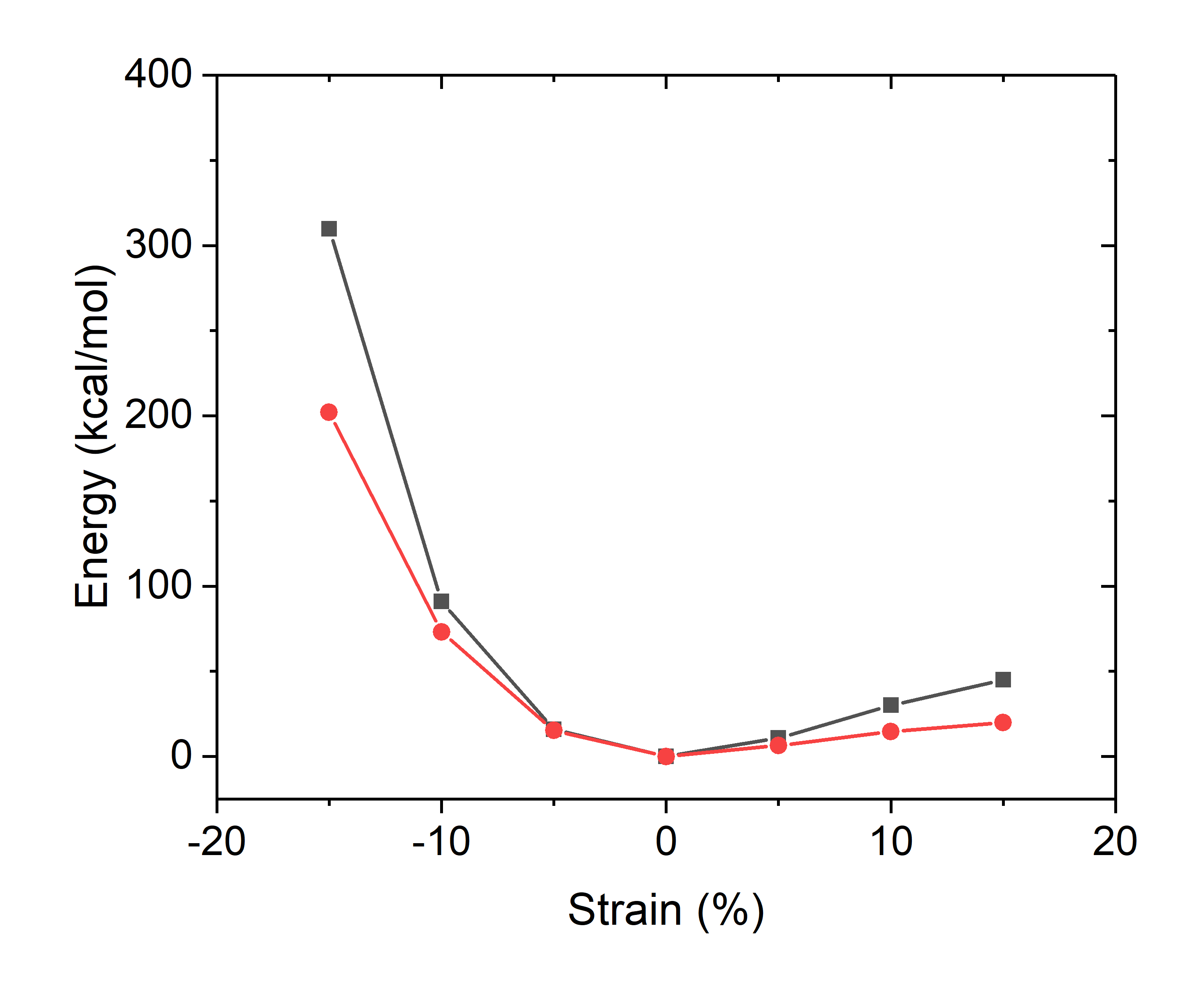}}
\caption{Octahedral}
\end{subfigure}
\begin{subfigure}[b]{.49\linewidth}
{\includegraphics[width=.95\textwidth]{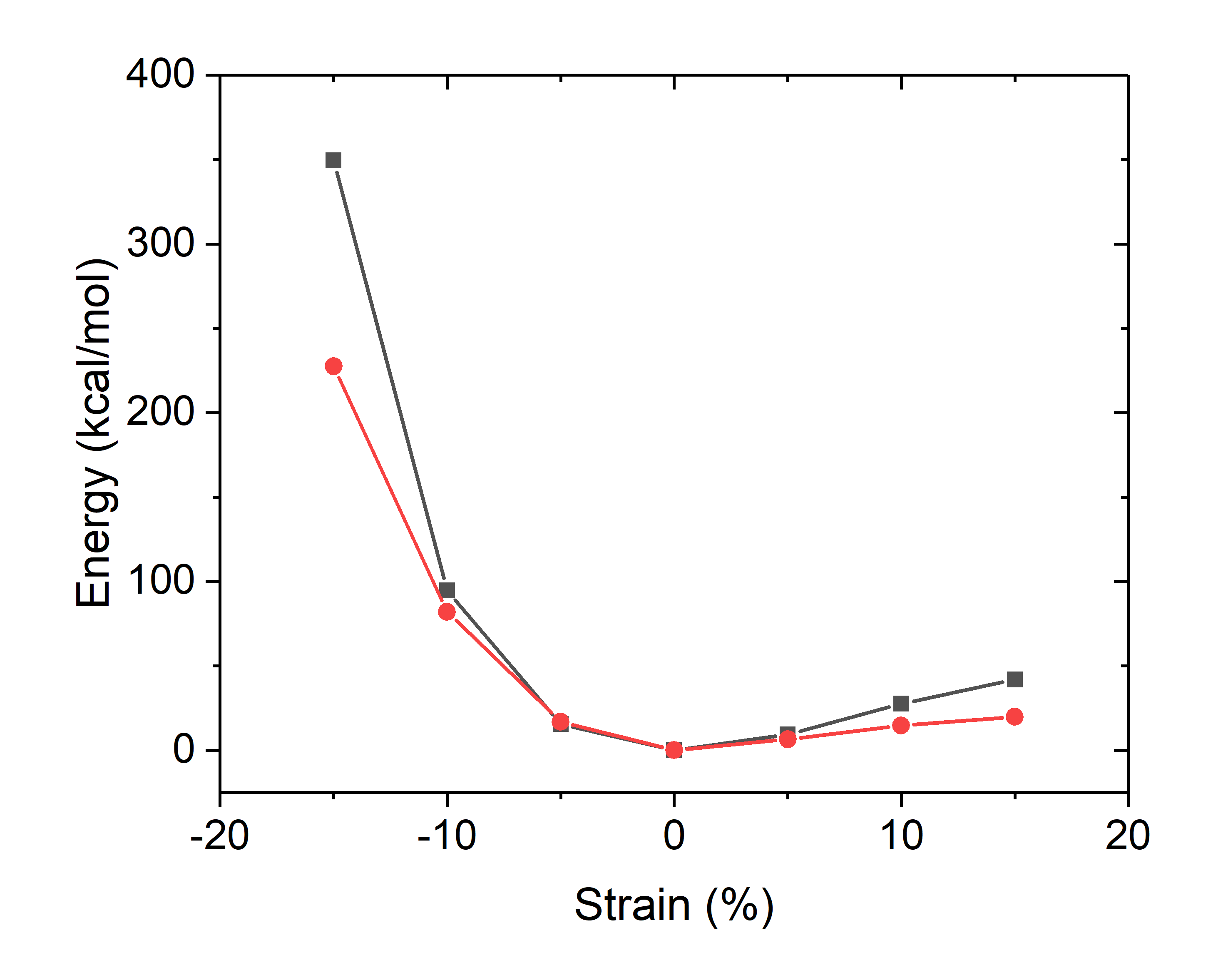}}
\caption{Tetrahedral}
\end{subfigure}
\caption{Equations of state calculated from DFT (red) and ReaxFF (black) for the Ni-doped $\mathrm{MoS_2}$ structures strained uniaxially in the $z$-direction. The inset in (b) shows a side view of the S-substituted structure with an arrow indicating the strain direction.}
\label{fig:EoS-uniZ}
\end{figure}

The energies of all four structures under all five strain conditions were used in the training of the ReaxFF force field.
The results for uniaxial straining in the $x$- and $z$-directions of the four Ni-doped $\mathrm{MoS_2}$ structures are shown in Figure~\ref{fig:EoS-uniX} and Figure~\ref{fig:EoS-uniZ}, respectively. 
The equation of state energies given are those of the doubled cell, with respect to the energy in each method of the unstrained structure. 
The structures used in ReaxFF result from a relaxation with fixed cell parameters, starting from the DFT structures.

We can begin by examining the energy difference between unstrained octahedrally and tetrahedrally intercalated structures. While Mo-substituted, S-substituted, and intercalated structures have different stoichiometries and cannot be compared in energy without assumptions about the chemical potential \cite{guerrero2020phase}, we can directly compare the energies of the two intercalated structures. 
In DFT, tetrahedral is lower than octahedral in energy by 38.8 kcal/mol in these structures (0.841 eV per Ni atom)\cite{guerrero2020phase}. 
We find a very close level of agreement in ReaxFF, which gives tetrahedral lower in energy by 41.0 kcal/mol. Note that in referring to energies in kcal/mol in this work we will consistently mean energies \textit{per unit cell}, which in most cases is a $2 \times 2$ cell. This agreement is essential for the correct structures to appear in MD simulations, and also demonstrates the ability of our ReaxFF parameterization to describe accurately the energy difference between the 6-bonded octahedral and 4-bonded tetrahedral environment.

The uniaxial ReaxFF energies are in reasonably good agreement with the DFT energies, despite the large strains that were applied.
Discrepancies are largest at large strain. Shapes are similar, though for $x$ strain and S-substituted, ReaxFF actually has a minimum shifted to +5\% strain (vs the DFT structure), and for $z$ strain and Mo-substituted the minimum is shifted to -5\%. 
For $x$ strain,  Mo-substituted has a larger energy value in ReaxFF than in DFT for all strains, whereas in the other cases, ReaxFF is higher for compressive strains and lower for tensile strains.
For $z$ strain, the ReaxFF energies are larger than in DFT in general, showing an overestimation of the elastic modulus in the $z$ direction. If we compare the level of discrepancy between ReaxFF and DFT here to the results of uniaxial strain for pristine MoS$_2$ in the 2017 parametrization (Fig. S5 in Ref. \cite{Ostadhossein2017}), we find the discrepancy is similar for the intercalated structures but larger for substituted structures.


\begin{figure}[H]
\centering
\begin{subfigure}[b]{.49\linewidth}
{\includegraphics[width=.95\textwidth]{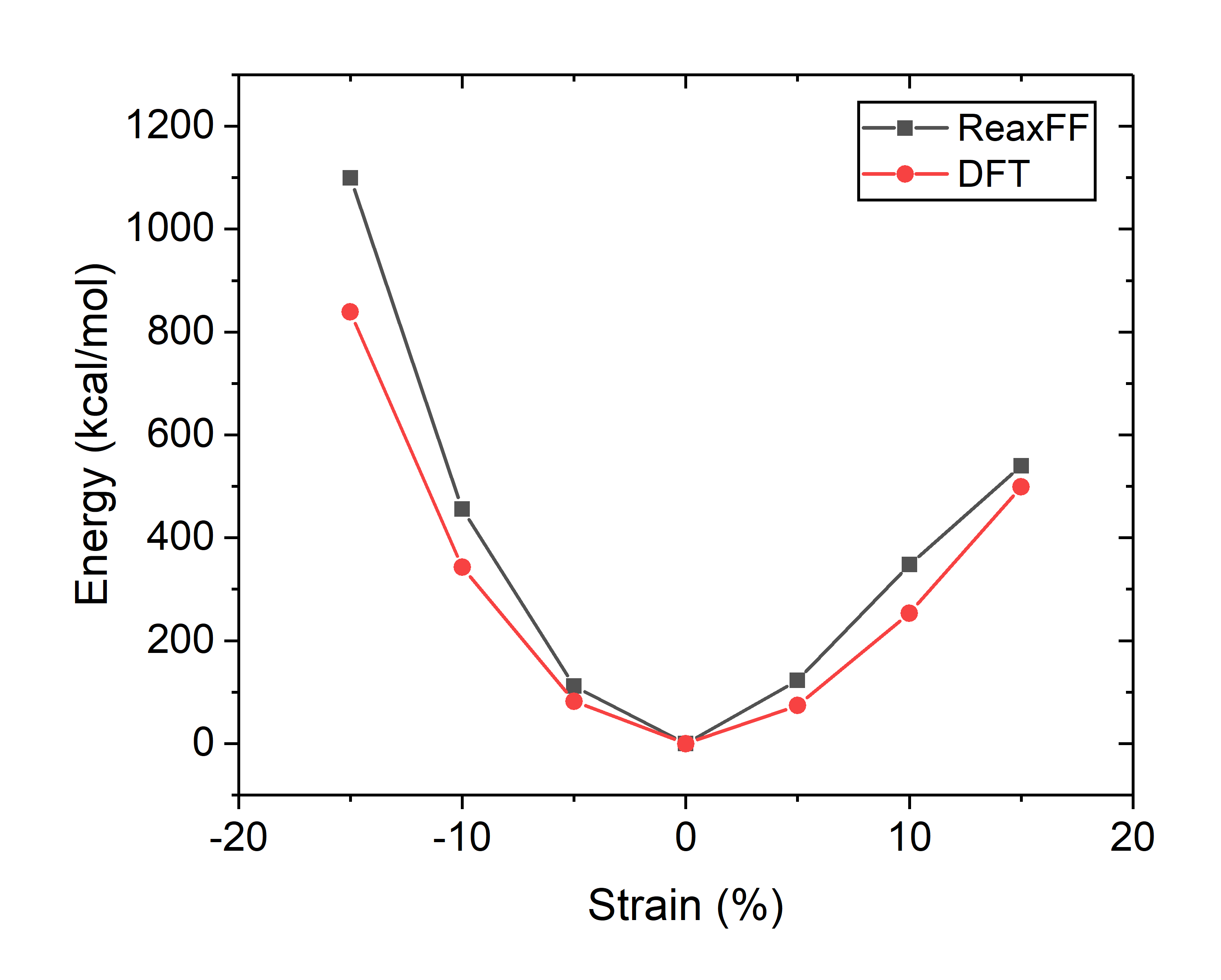}}
\caption{Mo-substituted}
\end{subfigure}
\begin{subfigure}[b]{.49\linewidth}
{\includegraphics[width=.98\textwidth]{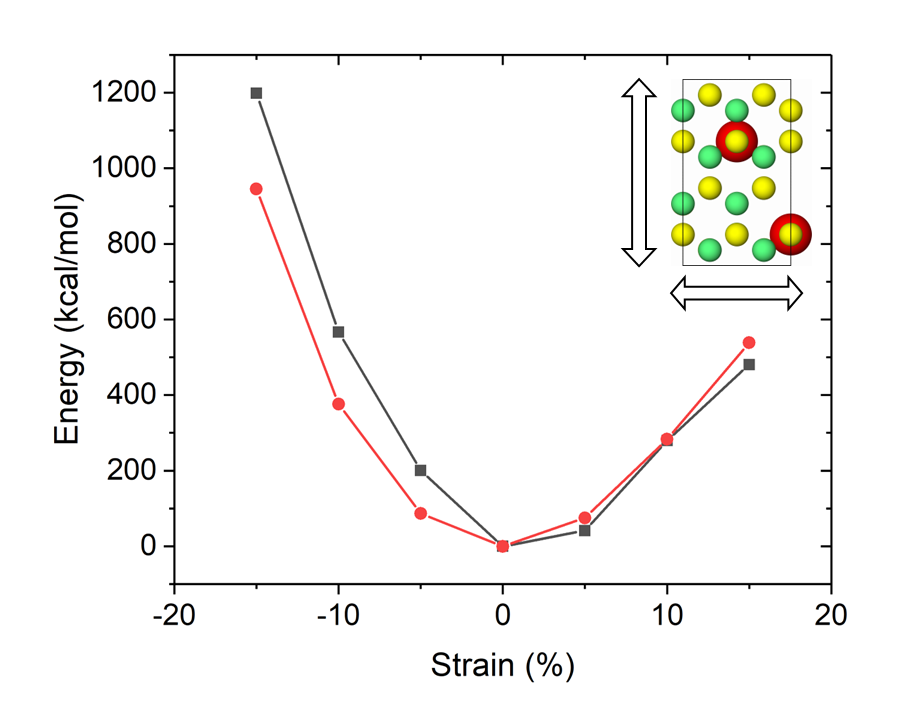}}
\caption{S-substituted}
\end{subfigure}
\begin{subfigure}[b]{.49\linewidth}
{\includegraphics[width=.95\textwidth]{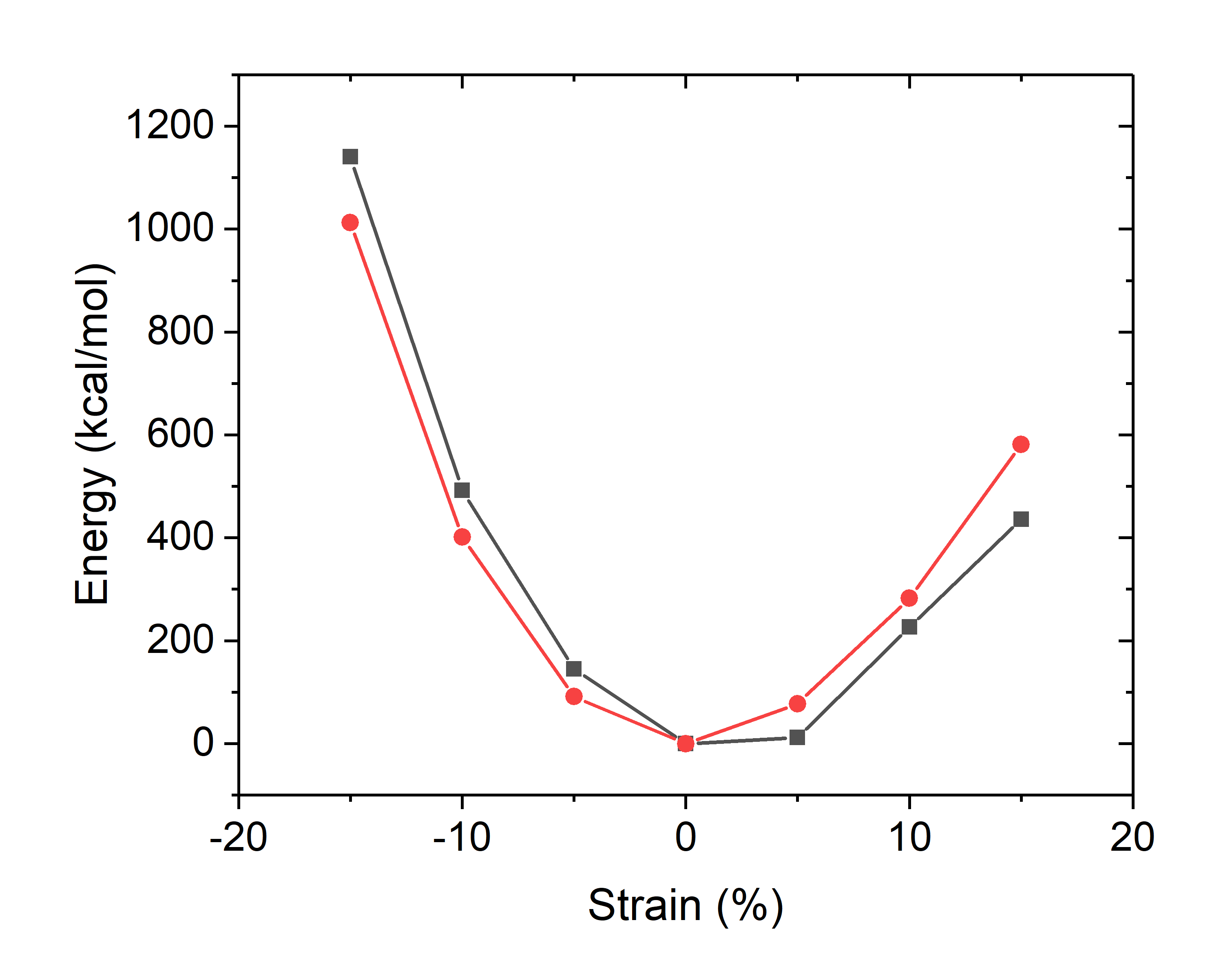}}
\caption{Octahedral}
\end{subfigure}
\begin{subfigure}[b]{.49\linewidth}
{\includegraphics[width=.95\textwidth]{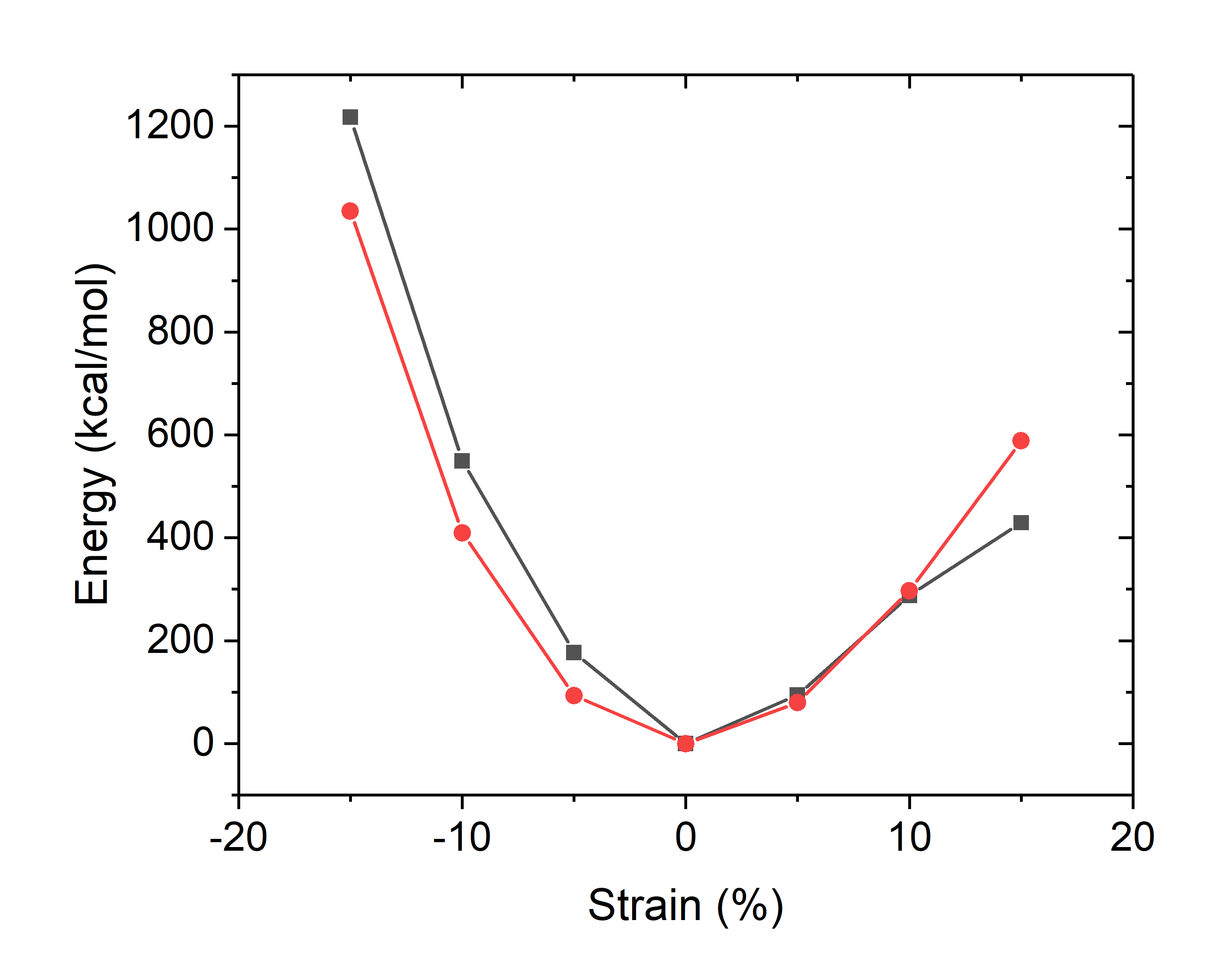}}
\caption{Tetrahedral}
\end{subfigure}
\caption{Equations of state as obtained from DFT (red) and ReaxFF (black) for the Ni-doped $\mathrm{MoS_2}$ structures strained biaxially for Mo-substituted, S-substituted, octahedral, and tetrahedral positions. The inset in (b) shows a top view of the S-substituted structure with two arrows indicating the strain directions.}
\label{fig:EoS-multi2}
\end{figure}

\begin{figure}
\centering
\begin{subfigure}[b]{.49\linewidth}
{\includegraphics[width=.95\textwidth]{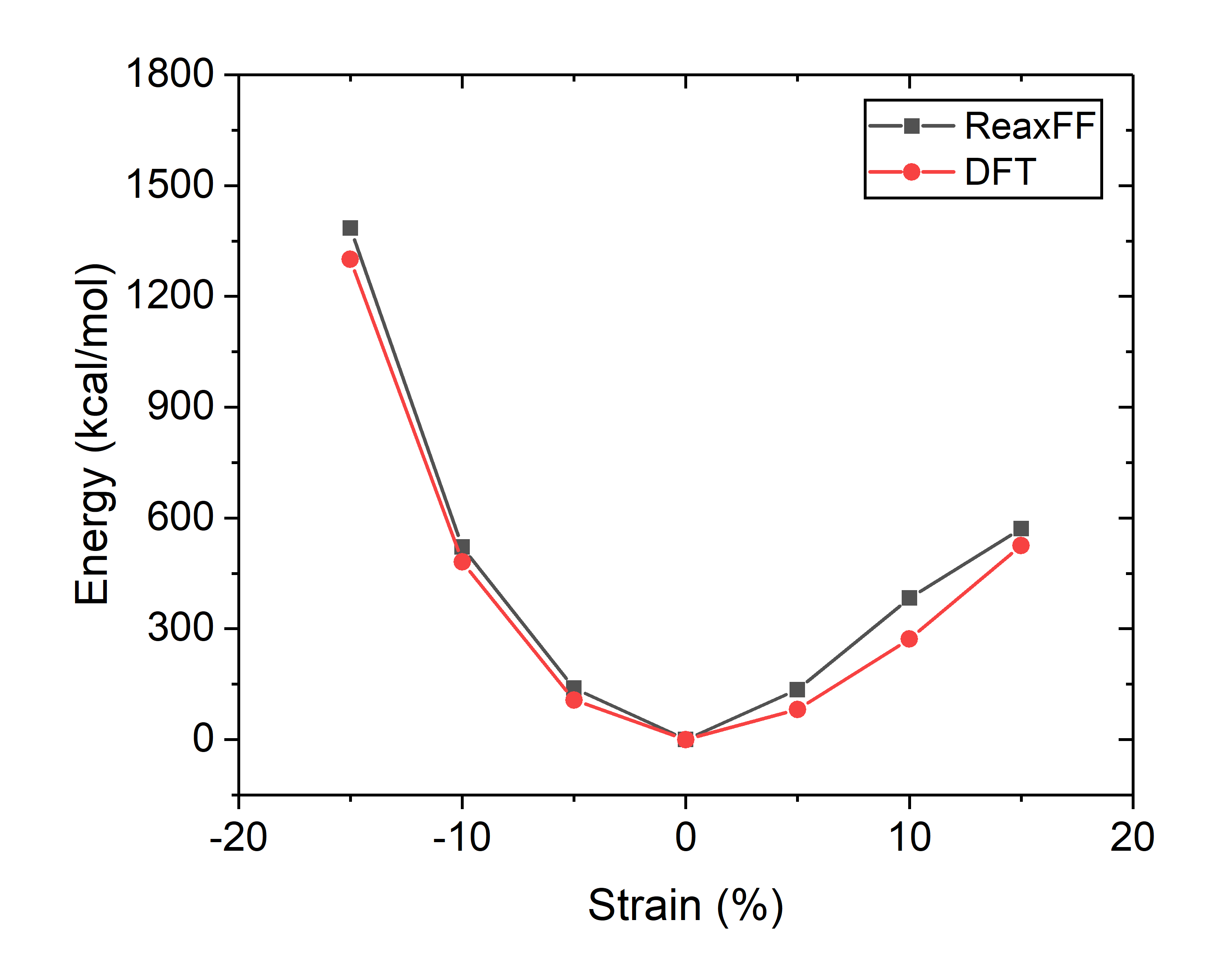}}
\caption{Mo-substituted}
\end{subfigure}
\begin{subfigure}[b]{.49\linewidth}
{\includegraphics[width=.98\textwidth]{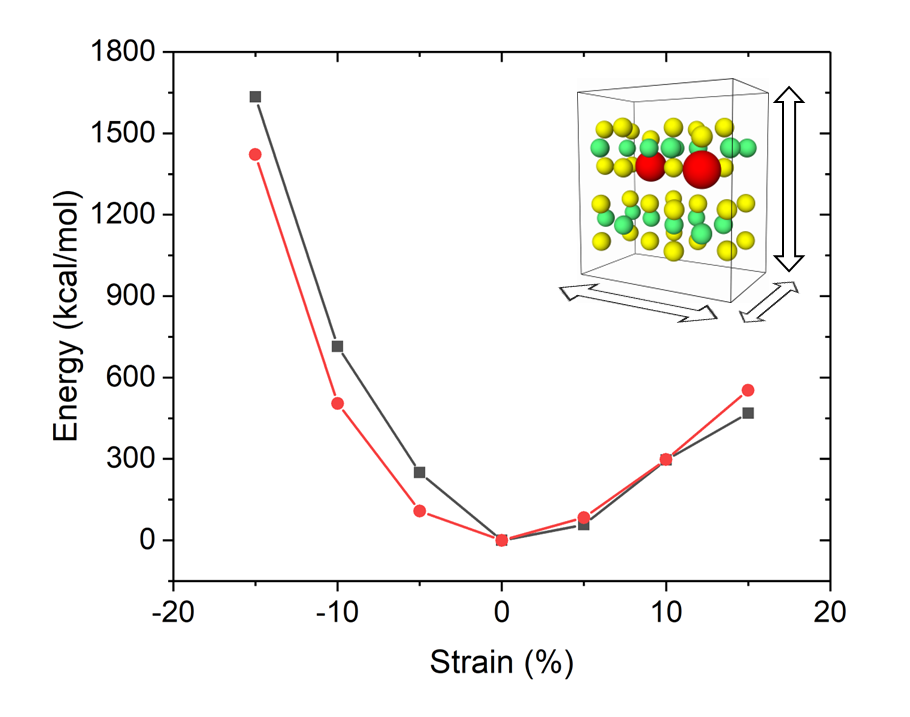}}
\caption{S-substituted}
\end{subfigure}
\begin{subfigure}[b]{.49\linewidth}
{\includegraphics[width=.95\textwidth]{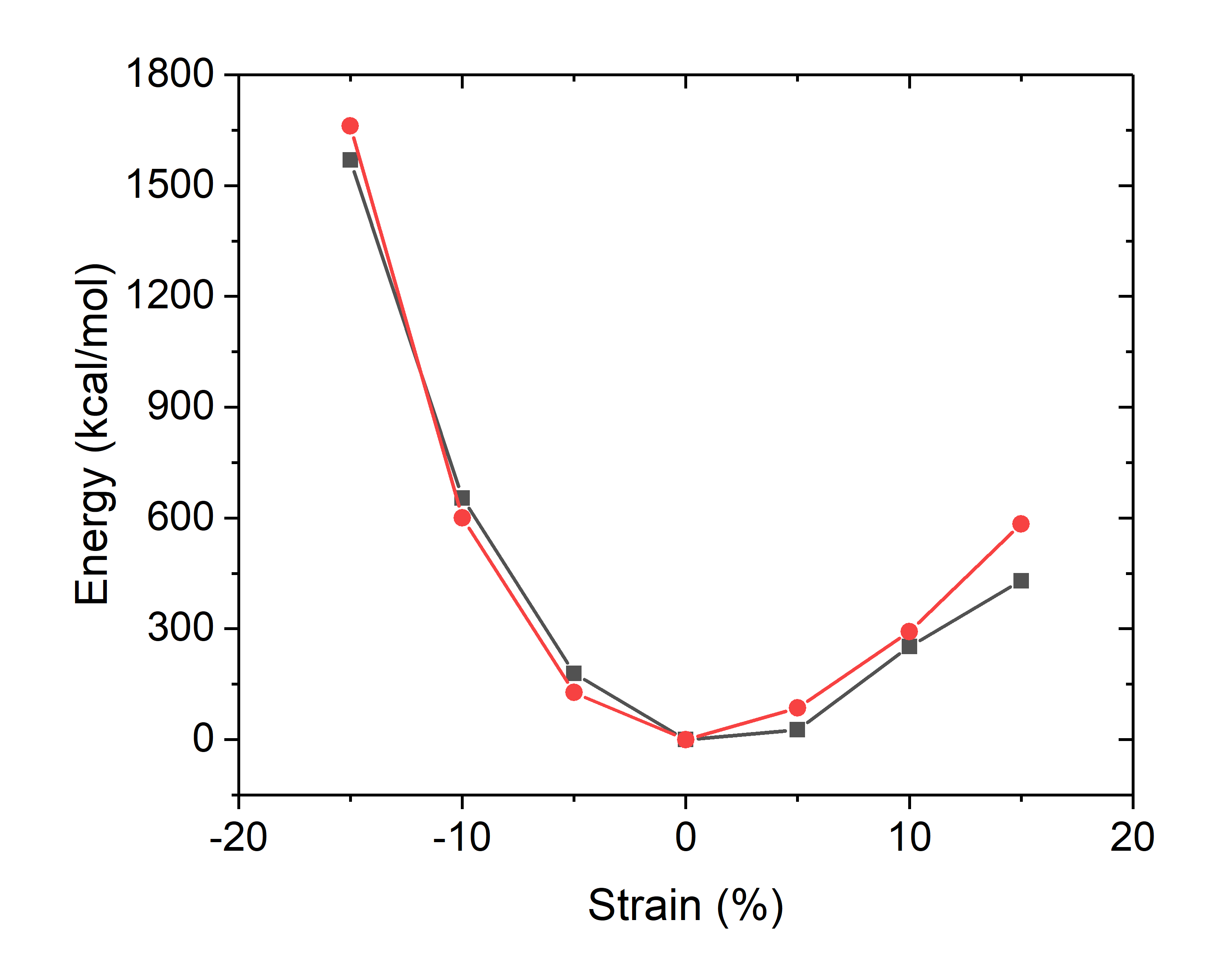}}
\caption{Octahedral}
\end{subfigure}
\begin{subfigure}[b]{.49\linewidth}
{\includegraphics[width=.95\textwidth]{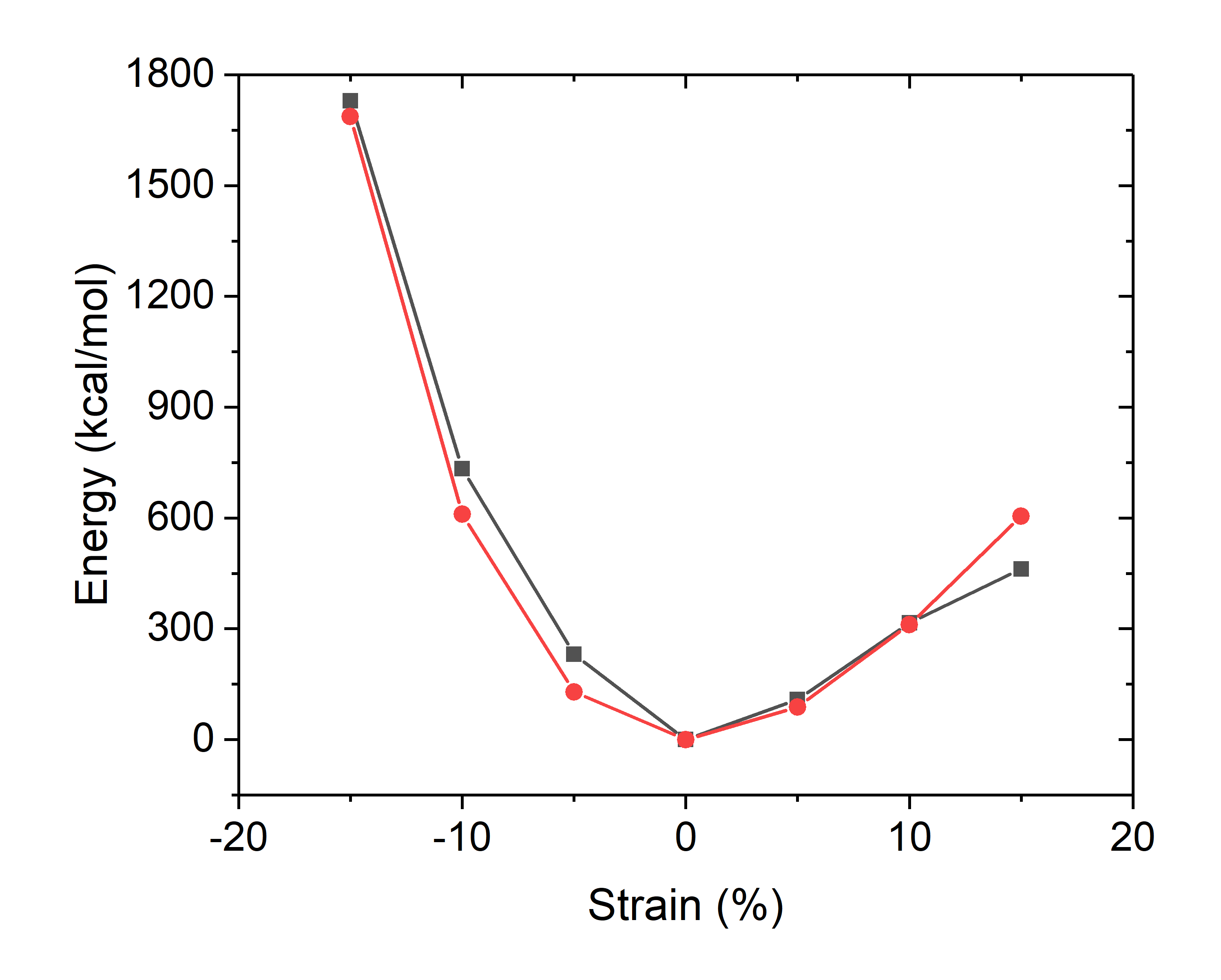}}
\caption{Tetrahedral}
\end{subfigure}
\caption{Equations of state as obtained from DFT (red) and ReaxFF (black) for the Ni-doped $\mathrm{MoS_2}$ structures strained triaxially for Mo-substituted, S-substituted, octahedral, and tetrahedral positions. The inset in (b) shows a perspective view of the S-substituted structure with three arrows indicating the strain directions.}
\label{fig:EoS-multi3}
\end{figure}

The results for biaxial and triaxial straining (where the specified strain value is applied to each axis) are shown in Figure~\ref{fig:EoS-multi2} and \ref{fig:EoS-multi3}, respectively, providing significantly better ReaxFF/DFT agreement than the uniaxial strain. We attribute the better agreement to the fact that bond angles change less in biaxial or triaxial strain than in uniaxial strain, so there is less dependence on the newly parametrized bond-angle terms in this work, and more dependence on the established Mo/S parameters we have taken from the literature.
The energies as obtained from DFT calculations for the highest strains are $\sim$~1000~kcal/mol for biaxial strain and even higher (up to $\sim$~1800~kcal/mol) in the case of triaxial strain.
Nevertheless, our force field is in good agreement with the DFT for the case of biaxial as well as triaxial straining.
The ReaxFF energies are consistent with DFT for both the near-equilibrium structures and the far-from-equilibrium energies, and the energy minimum is correctly at zero strain (unlike in Fig. \ref{fig:EoS-uniZ}(a)). 


\begin{figure}[H]
\begin{subfigure}[b]{.49\linewidth}
{\includegraphics[width=.95\textwidth]{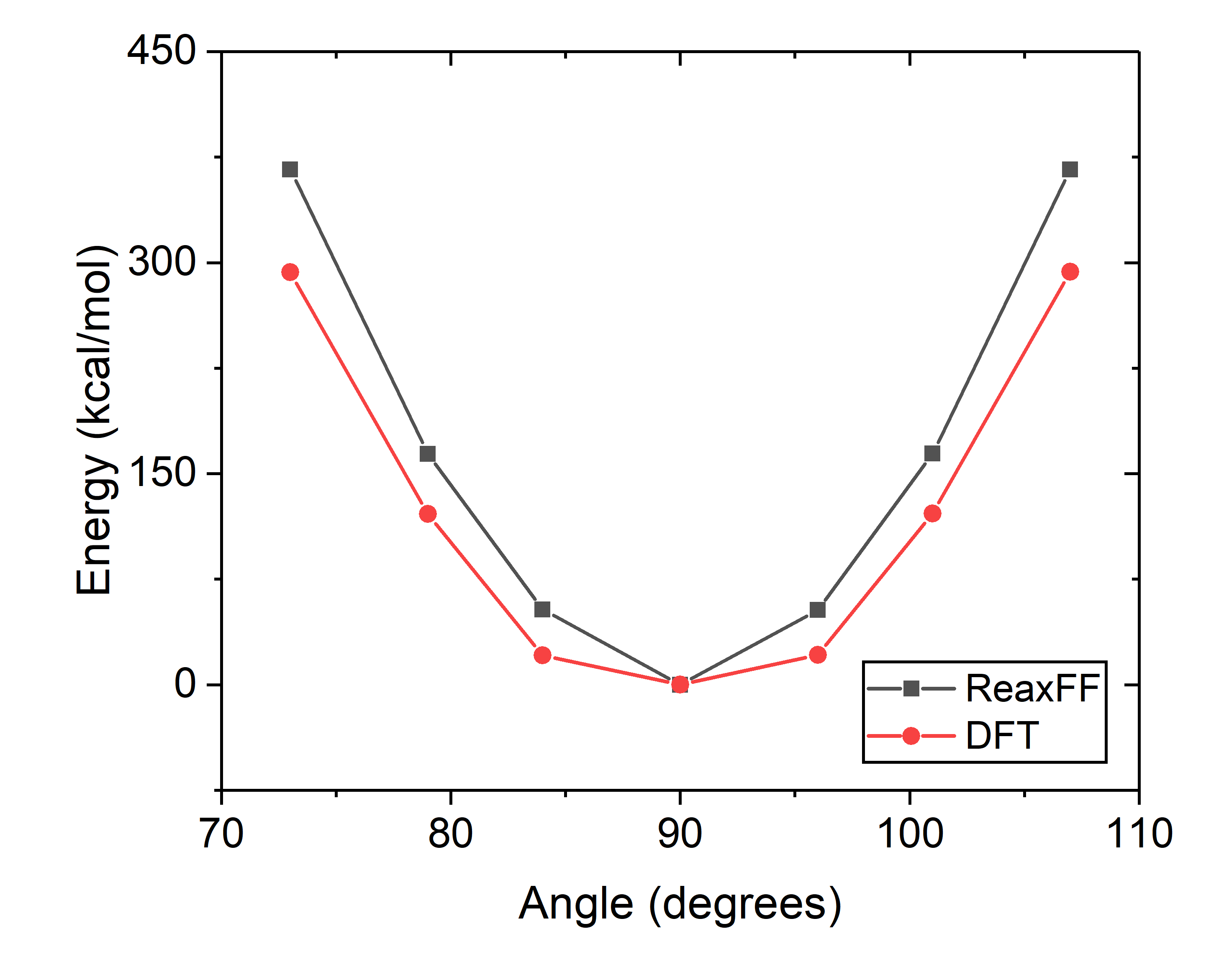}}
\caption{Mo-substituted}
\end{subfigure}
\begin{subfigure}[b]{.49\linewidth}
{\includegraphics[width=.98\textwidth]{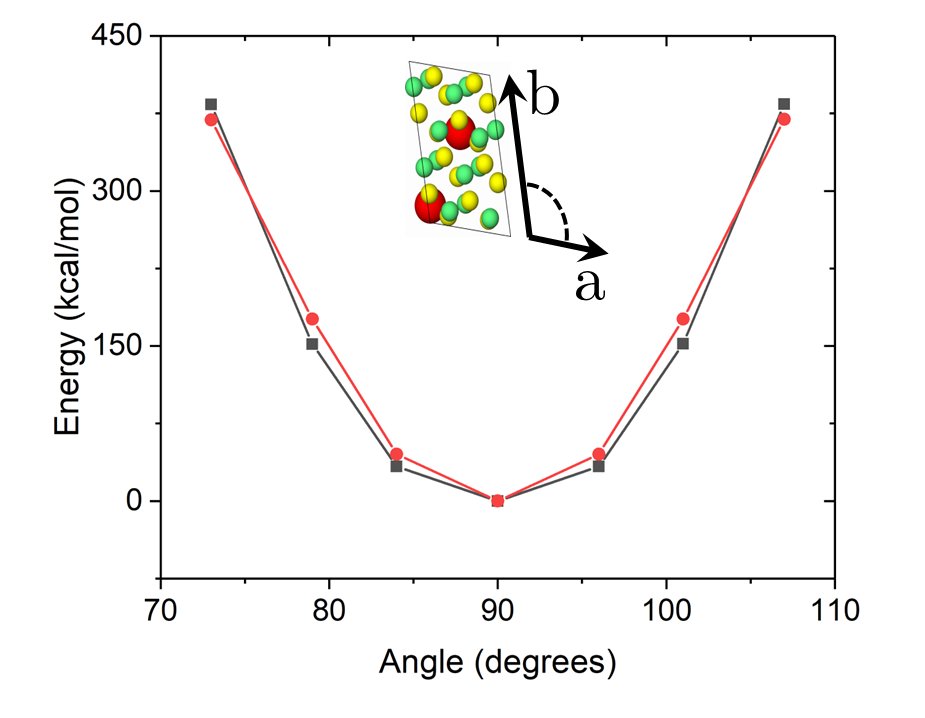}}
\caption{S-substituted}
\end{subfigure}
\begin{subfigure}[b]{.49\linewidth}
{\includegraphics[width=.95\textwidth]{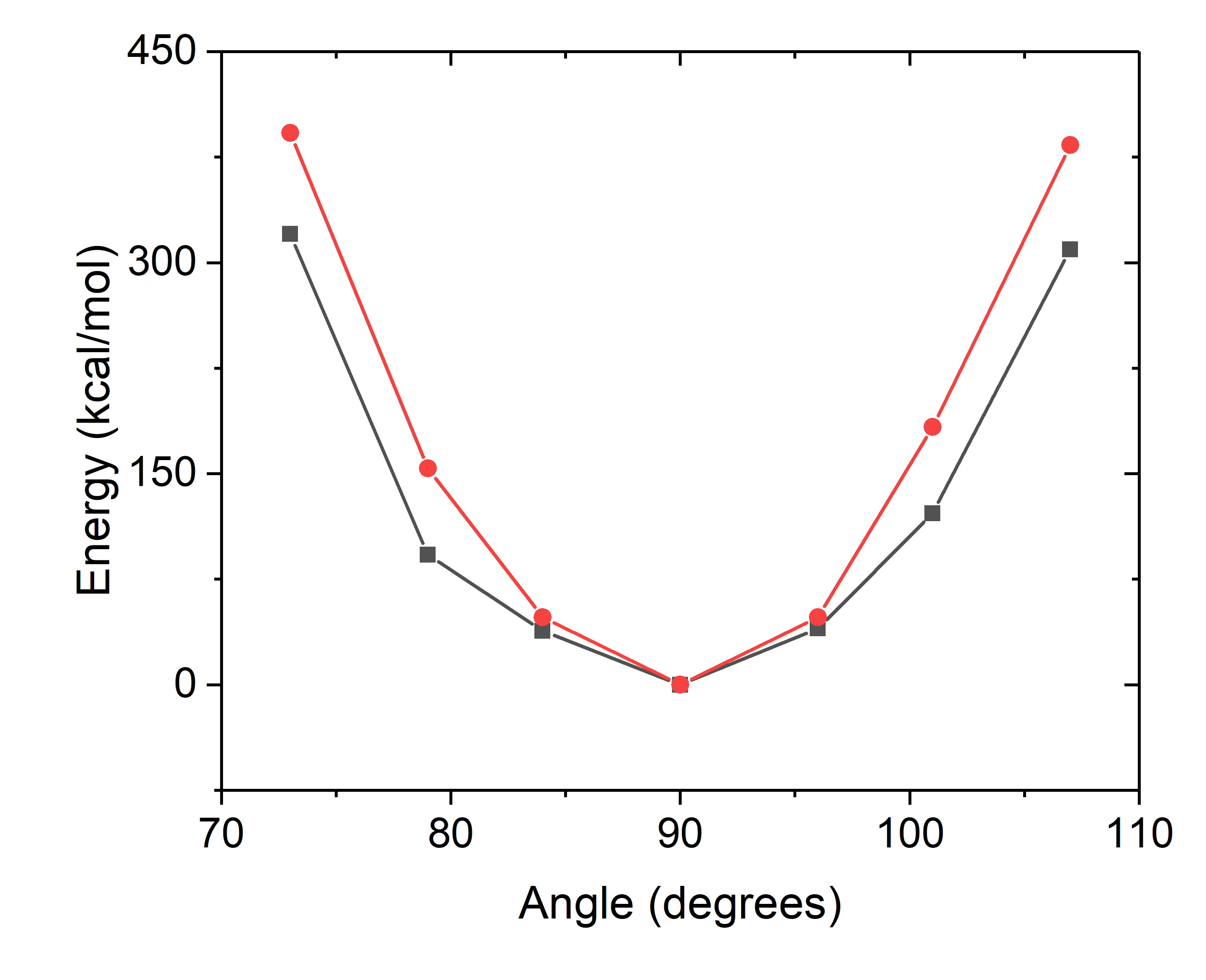}}
\caption{Octahedral}
\end{subfigure}
\begin{subfigure}[b]{.49\linewidth}
{\includegraphics[width=.95\textwidth]{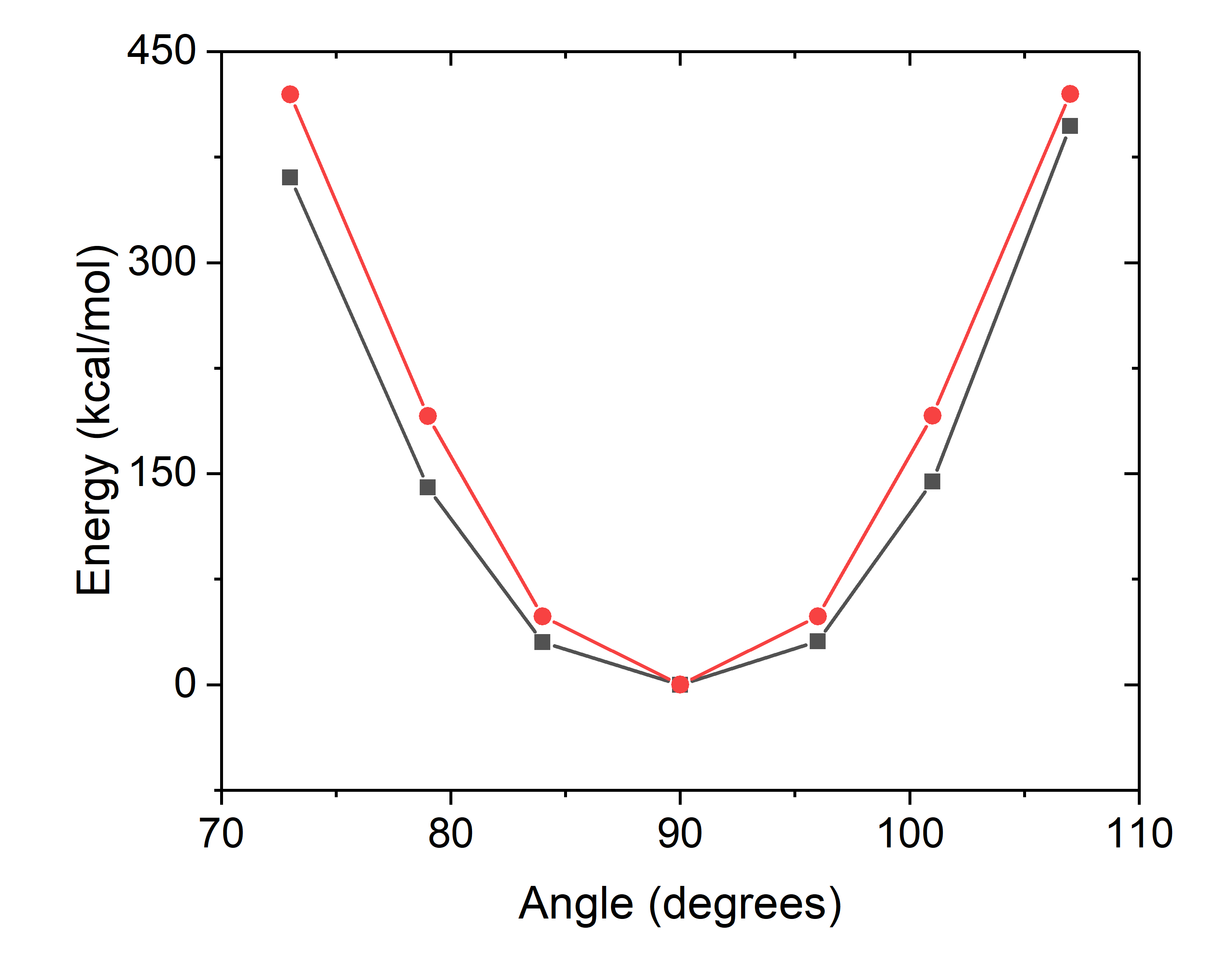}}
\caption{Tetrahedral}
\end{subfigure}
\caption{ReaxFF (black) and DFT (red) energies obtained for Ni-doped $\mathrm{MoS_2}$ structures sheared in the $xy$ basal plane for (a) Mo-substituted, (b) S-substituted, (c) octahedral, and (d) tetrahedral structures. The inset in (b) shows a top view of the basal plane of the S-substituted structure with arrows indicating the shearing angle as calculated from the angle between in-plane vectors $a$ and $b$ of the doubled cell.}
\label{fig:EoS-shear}
\end{figure}

Finally, the ReaxFF energies for sheared structures were compared with DFT. 
Figure~\ref{fig:EoS-shear} shows excellent agreement for near-equilibrium as well as far-from-equilibrium structures. 
The minimum is correctly at 90$^\circ$ and the shape is close and correctly symmetrical. Mo-substituted has an overestimated shear modulus whereas S-substituted is very close, and octahedral and tetrahedral intercalation have an underestimated shear modulus.

\subsection{Force Field Evaluation}
Next we evaluated the force field's ability to predict parameters that were not included in the training. 
This evaluation was based on atom positions and distances obtained from relaxation of the model structures using energy minimization with the conjugate gradient algorithm with force and energy criteria of $10^{-6}$~kcal/mol-\AA\ and $10^{-6}$~(unitless), respectively, and a simulation of periodic cell optimization with a target zero stress tensor. 
First, the Ni--Mo and Ni--S atomic distances in all four structures were calculated.
Results for Ni--Mo and Ni--S bond lengths are shown in Table~\ref{tab:bonds}a and \ref{tab:bonds}b, respectively.

\begin{table}[!htbp]
\caption{Atomic distances (below 3.6 \AA{}) between Ni and its neighbors from DFT and ReaxFF. Repeated distances are indicated with a multiplier. }
\flushleft
\begin{subtable}[c]{0.5\textwidth}
\begin{tabular}{||c | c | c||} 
 \hline
 Structure & DFT (\AA) & ReaxFF (\AA) \\ [0.5ex] 
 \hline\hline
 Mo-substituted & 3.20 $\times6$ & 3.19 $\times6$ \\ 
 \hline
 S-substituted & 2.55 $\times3$ & 2.76 $\times3$ \\ 
 \hline
 Octahedral & 3.57 $\times6$ & 3.30/3.44/3.47--4.36/4.38/4.50 \\ 
 \hline
 Tetrahedral  &  2.61 $\times1$  & 2.85 $\times1$\\ [1ex]
 \hline
\end{tabular}
\subcaption{Ni--Mo distances}
\end{subtable}
\begin{subtable}[c]{0.5\textwidth}
\centering
\begin{tabular}{||c | c | c||} 
 \hline
 Structure & DFT (\AA) & ReaxFF (\AA) \\ [0.5ex] 
 \hline\hline
 Mo-substituted & 2.38 $\times3$ & 2.36 $\times3$ \\ 
 \hline
 S-substituted & 3.18 $\times6$ & 3.25 $\times6$ \\ 
 \hline
 Octahedral & 2.34 $\times3$, 2.38 $\times3$ & 2.24/2.24/2.37--2.83/3.04/3.09 \\ 
 \hline
 Tetrahedral & 2.17 $\times3$, 2.12 $\times1$ & 2.32 $\times3$, 2.27 $\times1$\\ [1ex]
 \hline
\end{tabular}
\subcaption{Ni--S distances}
\end{subtable}
\label{tab:bonds}
\end{table} 

\begin{figure}[H]
\centering
{\includegraphics[width=0.95\textwidth]{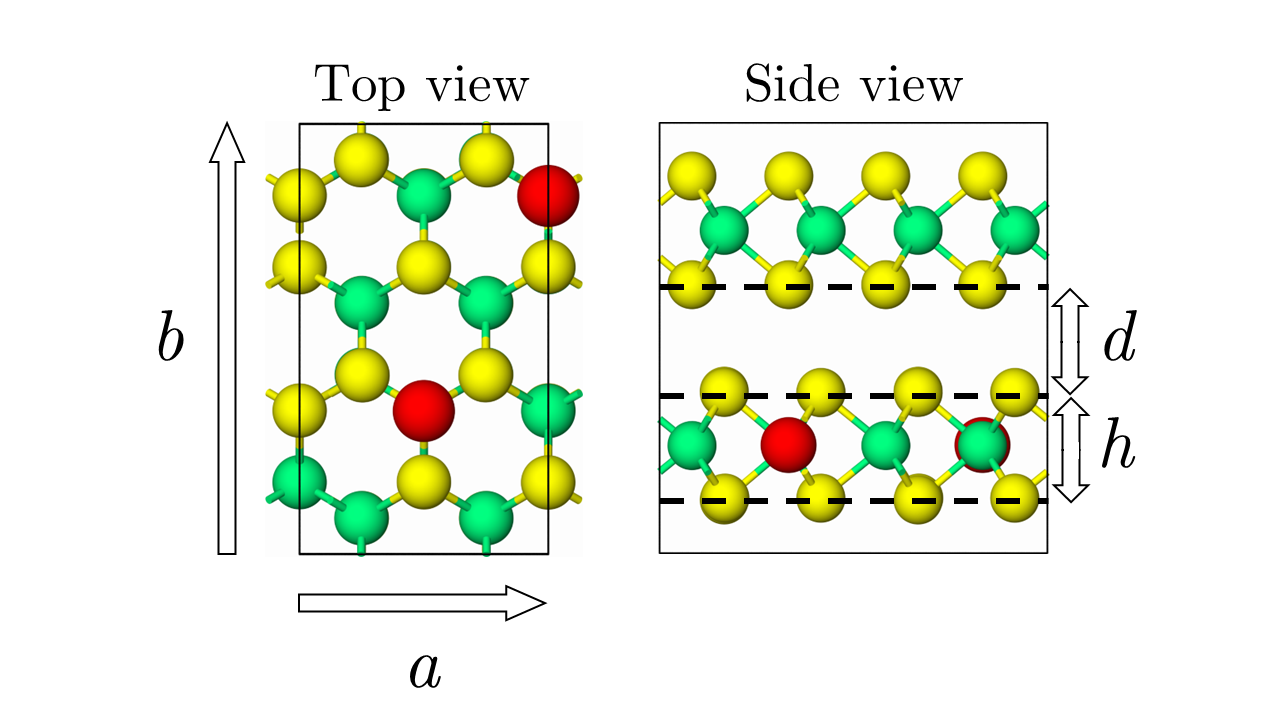}}\\
\caption{Structural parameters for Ni-doped $\mathrm{MoS_2}$ illustrated for Mo-substituted configuration: cell vectors $a$ and $b$, average distance $h$ between S planes in a single layer, and average interlayer separation $d$.}
\label{fig:Lattice}
\end{figure}

In most cases, the differences between ReaxFF and DFT bond lengths were within 0.1~\AA, indicating the force field can accurately capture bond lengths within the Ni-doped $\mathrm{MoS_2}$ structure.
The same analysis for Mo--S bond lengths revealed that the difference between the DFT and ReaxFF calculated values was less than 0.1~\AA{} for most cases and less than 0.2~\AA{} for some cases in the octahedral and tetrahedral intercalation structures.
For the pristine structure, the Mo--S bond length difference between DFT and ReaxFF calculated using the original 2022 parameters was 0.04\AA{}.
Generally, for Ni-doped $\mathrm{MoS_2}$, ReaxFF predicted a slightly larger interlayer separation resulting an increase in the bond distances.
The local symmetry around Ni, as shown by the multiplicity of the atomic distances, was correctly preserved in every case except the octahedral intercalation. 
However, for octahedral intercalation, the larger separation of the two $\mathrm{MoS_2}$ layers in ReaxFF resulted in breaking of the symmetry that was predicted by DFT calculations, and much larger deviations in bond lengths.
Next, the force field's ability to reproduce Ni-doped $\mathrm{MoS_2}$ structures with correct lattice parameters was tested. 
The parameters considered, illustrated in Figure~\ref{fig:Lattice}, were the doubled cell vectors, $a$ and $b$, the average distance $h$ between S planes in a single layer, and the average interlayer separation $d$.
The values for these parameters after structure relaxation in DFT and in ReaxFF are shown in Table~\ref{tab:LatticeParams}.
The $d$ parameter decreases with doping for all structures in both ReaxFF and DFT, which is consistent with an experimentally measured reduction in $c$-parameter.\cite{Garreau1986} 
In most experiments, however, the atomistic structure of the Ni-doped MoS$_2$ is unknown, making precise comparisons between experiment and DFT unclear.
The difference between the DFT- and ReaxFF-calculated parameters was less than 0.1~\AA{} for most doped structures, as well as the pristine $\mathrm{MoS_2}$ case which is shown for reference and relies only on the pre-existing Mo-S potential.
This is consistent with the equation of state findings, since the optimized values of $a$ and $c = 2h + 2d$ are the minima of the uniaxial $x$ and $z$ curves.
The most notable discrepancy is for $d$ in the octahedral case, where ReaxFF overestimates by $\sim$~0.4~\AA.

\begin{center}
\begin{table}[]
\caption{Comparison of structural parameters as obtained from DFT and ReaxFF for bulk Ni-doped $\mathrm{MoS_2}$}
\begin{tabular}{|l|cc|cc|cc|cc|}
\hline
\multicolumn{1}{|c|}{\multirow{2}{*}{Structure}} & \multicolumn{2}{c|}{$a$~(\AA)}             & \multicolumn{2}{c|}{$b$~(\AA)}              & \multicolumn{2}{c|}{$d$~(\AA)}             & \multicolumn{2}{c|}{$h$~(\AA)}             \\ \cline{2-9} 
\multicolumn{1}{|c|}{}                           & \multicolumn{1}{c|}{DFT}  & ReaxFF & \multicolumn{1}{c|}{DFT}   & ReaxFF & \multicolumn{1}{c|}{DFT}  & ReaxFF & \multicolumn{1}{c|}{DFT}  & ReaxFF \\ \hline
Pristine                                         & \multicolumn{1}{c|}{6.38} & 6.40   & \multicolumn{1}{c|}{11.04} & 11.10  & \multicolumn{1}{c|}{3.08} & 3.03   & \multicolumn{1}{c|}{3.12} & 3.21   \\ \hline
Mo-substituted                                   & \multicolumn{1}{c|}{6.40} & 6.40   & \multicolumn{1}{c|}{11.08} & 11.08  & \multicolumn{1}{c|}{3.07} & 3.00   & \multicolumn{1}{c|}{3.07} & 3.16   \\ \hline
S-substituted                                    & \multicolumn{1}{c|}{6.35} & 6.47   & \multicolumn{1}{c|}{11.00} & 11.21  & \multicolumn{1}{c|}{2.87} & 2.77   & \multicolumn{1}{c|}{3.15} & 3.30   \\ \hline
Octahedral                                       & \multicolumn{1}{c|}{6.38} & 6.45   & \multicolumn{1}{c|}{11.06} & 11.20  & \multicolumn{1}{c|}{2.66} & 3.03   & \multicolumn{1}{c|}{3.12} & 3.17   \\ \hline
Tetrahedral                                      & \multicolumn{1}{c|}{6.39} & 6.43   & \multicolumn{1}{c|}{11.07} & 11.16  & \multicolumn{1}{c|}{2.88} & 2.95   & \multicolumn{1}{c|}{3.06} & 3.14   \\ \hline
\end{tabular}
\label{tab:LatticeParams}
\end{table}

\end{center}

\begin{figure}[H]
\centering
{\includegraphics[width=0.95\textwidth]{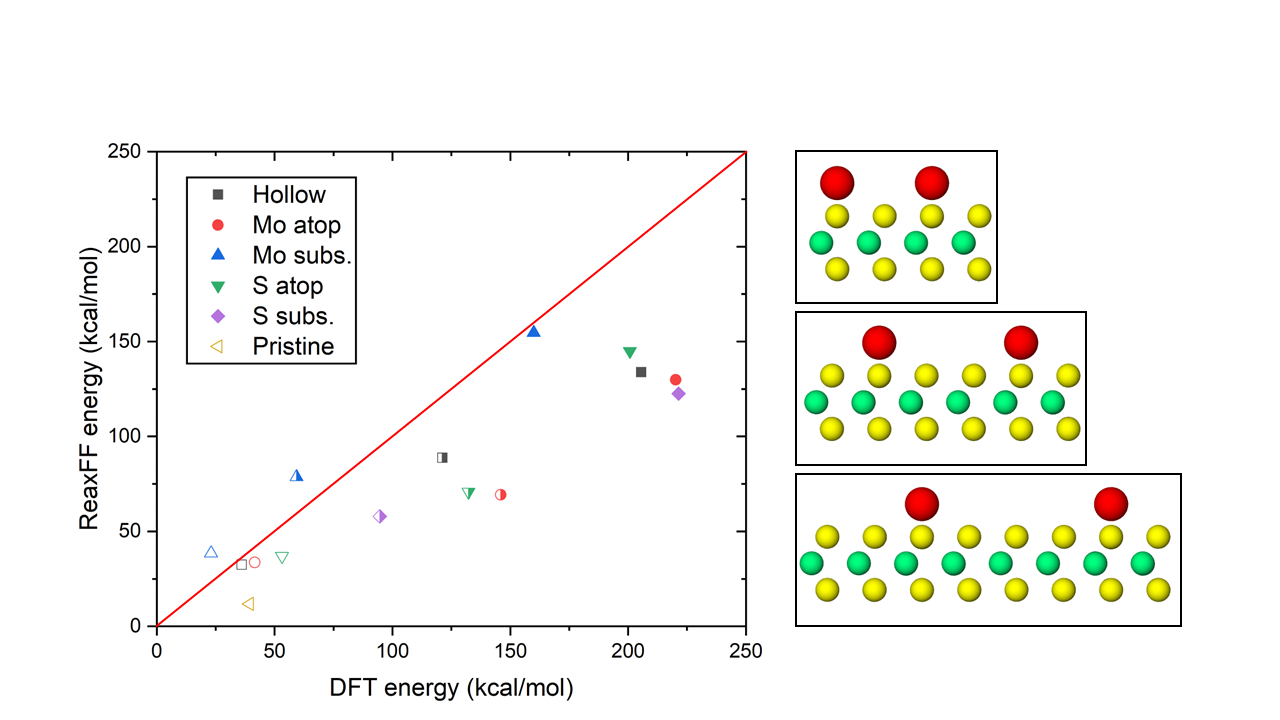}}\\
\caption{Energy difference between 1H and 1T polytypes for each structure of Ni-doped (or pristine) $\mathrm{MoS_2}$. 
Results show energy differences of the right scale for ReaxFF, but with less variation, leading to underestimation by ReaxFF for some doping locations. 
Change in the concentration from high to low is illustrated with open, half-filled, and solid symbols for $2\times2$, $3\times3$, and $4\times4$ supercells, respectively.
Snapshots (right) show $2\times2$, $3\times3$, and $4\times4$ supercells of S atop site of 1H-$\mathrm{MoS_2}$ (different size supercells correspond to different doping concentrations).}
\label{fig:Poly_size}
\end{figure}

The accuracy of the developed force field was also tested in distinguishing the relative energies of Ni-doped 1H and 1T monolayer  structures.
We used 2$\times$2, 3$\times$3, and 4$\times$4 in-plane supercells of the three-atom unit cell of 1H and 1T, where each supercell contained one Ni atom; these structures were then doubled to create orthogonal unit cells with two Ni atoms per cell. 
The different doping sites~\cite{karkee2021structural,Karkee_panoply} were: adatoms at the hollow position (three-fold hollow space between top S atoms), Mo atop (on top of Mo), or S atop (on top of S); Mo-substituted, or S-substituted.
The pristine structures were also included for reference, which had been studied in the 2017 work.\cite{Ostadhossein2017}
Note that neither 1H or 1T structures nor adatoms were in our training data. 
Snapshots of representative structures (S atop) of different sizes (doping concentrations) are shown in Figure~\ref{fig:Poly_size}. 
The energy differences between 1H and 1T polytypes were compared between ReaxFF and DFT in Figure~\ref{fig:Poly_size}, using DFT results from \citet{karkee2021structural} and \citet{Karkee_panoply}.
Results show that the force field can capture the energies corresponding to Ni dopants and adatoms with a small underestimation, as in the pristine case.\cite{Ostadhossein2017}.

Given the prevalence of vacancies in real $\mathrm{MoS_2}$~\cite{zhou2013intrinsic} samples, and studies of vacancies with the 2017 ReaxFF potential,\cite{Ostadhossein2017} we tested our reactive force field on defective bulk 2H structures that combine Mo or S vacancies and Ni dopants. In each case, the vacancy was located at the nearest Mo or S site to the Ni dopant. 
Structures based on a $3 \times 3 \times 1$ supercell were constructed and underwent variable-cell relaxation in DFT, and then the DFT-relaxed structure was converted to a nearly orthorhombic doubled cell, imported to ReaxFF and relaxed using the developed force field.
The goal to have an agreement between DFT and ReaxFF required that the final (i.e., after relaxation) atomic positions of Ni-doped $\mathrm{MoS_2}$ atoms be the same (i.e., no structural changes during ReaxFF relaxation). 
A summary of atomic rearrangements during each relaxation is shown in Table~\ref{tab:vacancy}.

\begin{center}
\begin{table}
 \begin{tabular}{|| c | c | c | c ||} 
 \hline
 Ni initial site & Vacancy initial site & Ni relaxed site & Vacancy relaxed site\\ [0.5ex] 
  \hline\hline
  Mo & Mo & Mo & Mo \\ 
  \hline
  Mo & S & Mo & S \\ 
  \hline
  S & Mo & Mo & S \\ 
  \hline
  S & S & S & S \\
  \hline
  Octahedral & Mo & Mo & - \\ 
  \hline
  Octahedral & S & S & - \\ 
  \hline
  Tetrahedral & Mo & Mo  & - \\ 
  \hline
  Tetrahedral & S & S  & - \\ [1ex]
  \hline
\end{tabular}
\caption{\label{tab:vacancy}{Summary of atomic rearrangements of 2H-MoS$_2$ with a Ni dopant and a vacancy, both at variable locations. In intercalations and S-substituted with a vacancy, Ni moves into the vacancy. All rearrangements predicted by DFT remain in ReaxFF. }}
\end{table}
\end{center}

\begin{figure}[H]
\centering
\begin{subfigure}[b]{.49\linewidth}
{\includegraphics[width=.95\textwidth]{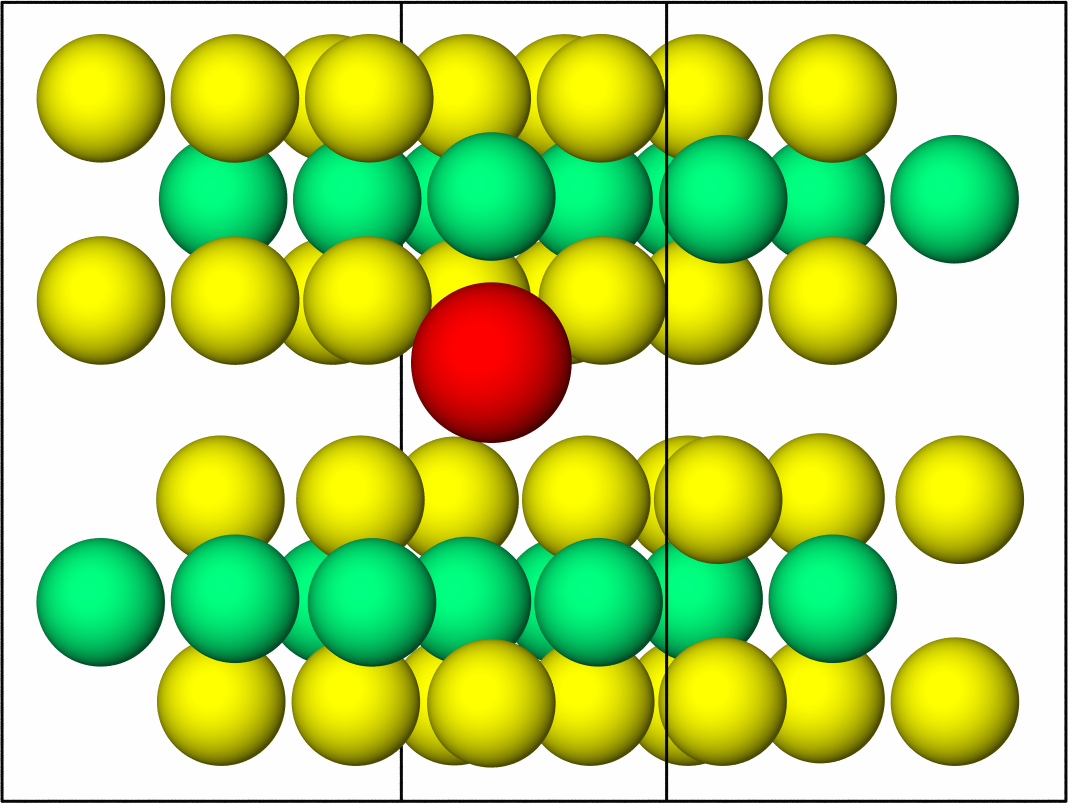}}
\caption{Before}
\end{subfigure}
\begin{subfigure}[b]{.49\linewidth}
{\includegraphics[width=.97\textwidth]{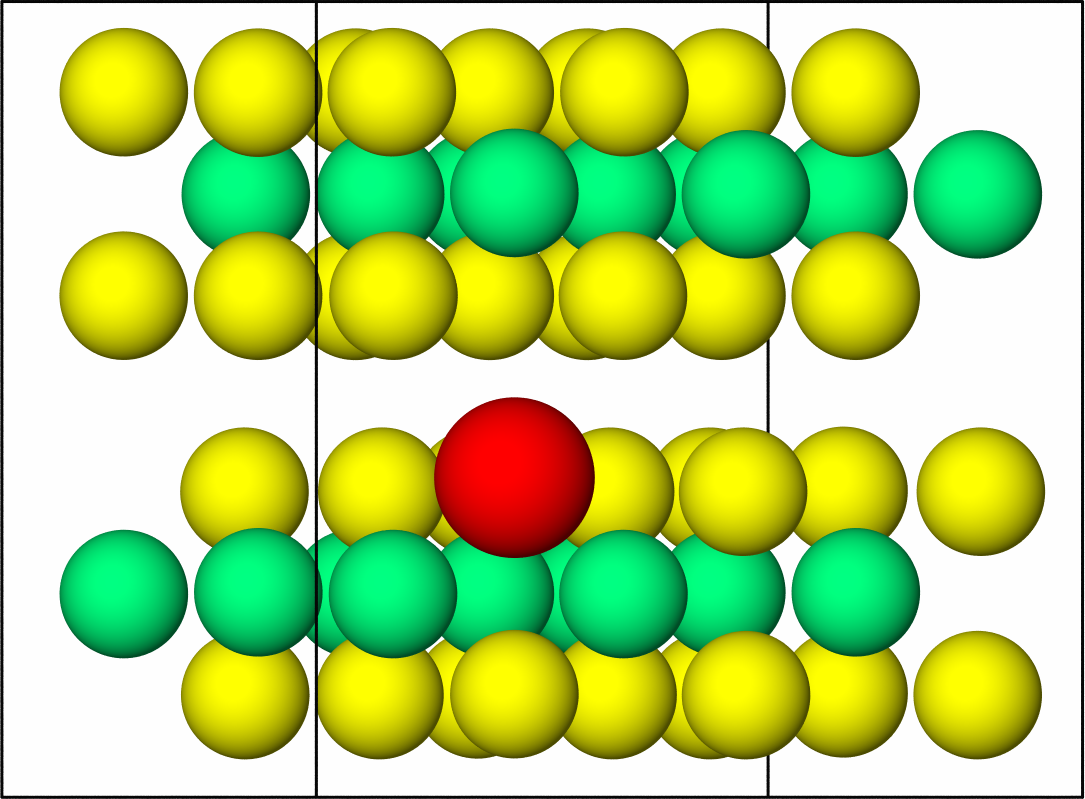}}
\caption{After}
\end{subfigure}
\caption{Relaxation of a dopant/vacancy structure as calculated by DFT: (a) tetrahedrally intercalated Ni adjacent to an S vacancy; (b) Ni has moved into the S site, forming S-substituted MoS$_2$.}
\label{fig:vac}
\end{figure}

Table~\ref{tab:vacancy} shows that, according to DFT, for the intercalated structures, Ni moved to the vacancy position, essentially converting an intercalated structure to a substituted structure (Mo-substituted for Ni filling an Mo vacancy and S-substituted for Ni filling an S vacancy). An example is shown in Fig. \ref{fig:vac}. This observation is consistent with the findings in \citet{guerrero2020phase} that the formation energy for tetrahedral intercalation in a $3 \times 3 \times 1$ supercell is 0.401 eV, greater than the energy for filling an Mo vacancy with Ni (-2.575 eV), and for filling an S vacancy (0.194 eV). The formation energy for octahedral intercalation is 0.9 eV higher, making the migration to a vacancy yet more exothermic. Since relaxation led to the migrations, evidently these migrations have no barrier. We additionally observe that S-substituted Ni adjacent to an Mo vacancy moves to the Mo site, leaving behind an S vacancy. 
The ReaxFF relaxed structures preserved the same rearrangements as in DFT for all combinations of dopant and vacancy positions.

Lastly, we evaluated the potential in terms of its ability to model interlayer sliding. Here, we rigidly displaced layers in the $x$- or $y$-directions, then relaxed in the $z$-direction, as in \citet{Guerrero_sliding}.
However, as shown in Figs. S10-15, the sliding energies from ReaxFF did not match those from DFT, except for the Mo-substituted case sliding in the $x$-direction. In many cases, sliding leads to a reduction in energy in ReaxFF, i.e. an incorrect lowest-energy stacking is predicted vs. DFT.
Even the original potentials could not capture sliding energies accurately (Figs. S12 and S15), since neither the original nor the doped potential was trained for sliding. 

\subsection{Simulations of Deposition and Annealing}

To demonstrate the applicability of the newly developed force field, the process of sputter deposition and annealing to grow Ni-doped $\mathrm{MoS_2}$ films~\cite{sirota2019MoS2-sputtering-exp} was simulated.
The simulations were performed using LAMMPS in the NVT ensemble with the Langevin thermostat, a damping parameter of 10.0~fs, and a time step of 0.1~fs.
The simulation box was $2.5\times2.4\times10.0$~nm in the $x$-, $y$-, and $z$-directions, respectively, with periodic boundary conditions in the $x$- and $y$-directions.
The boundary condition in the $z$-direction was fixed to mimic deposition in this direction.
The initial configuration was a substrate consisting of bilayer 2H-$\mathrm{MoS_2}$, with the bottommost layer held fixed during the simulation.

Atoms were deposited from 7.0~nm above the substrate surface, and a reflective virtual wall (which reflects downward only) was placed parallel to the surface at a distance of 4.8~nm to ensure deposited atoms remain near the substrate.  
The deposition process followed a simulation protocol used previously for $\mathrm{SiO_2}$ thin film formation.~\cite{Taguchi2007-cv} While the timescales accessible in atomistic simulations are much shorter than in real experiments, this kind of simulation has been found to be useful in generating realistic atomistic structures.
First, energy minimization was performed to obtain the relaxed atomic positions, followed by thermal equilibration for 50~ps at room temperature.
Next, Mo, S, and Ni atoms were continuously deposited onto the $\mathrm{MoS_2}$ substrate at a 1:2 Mo to S ratio with Ni atoms replacing Mo atoms at 7\% by weight, expected to promote Ni substitution for Mo.~\cite{Azhar2020Nidoped} 
The deposition rates for Mo and Ni were one atom every 40~fs with a deposition energy of 230~kcal/mol; for S atoms the deposition rate was one atoms every 20~fs with a deposition energy of 1.5~kcal/mol. Under typical experimental conditions, an inert gas such as argon is used to modulate the incident kinetics of the deposited Mo, S, and Ni radicals. Following previous simulations of Si-O deposition~\cite{Grigoriev2015-wg}, a lower deposition energy was used to slow the sulfur atoms such that they act not only as radicals but as kinetic energy modulators, without the need to explicitly model argon.
The total number of deposited S, Mo, and Ni atoms after 50~ps was 1000, 405, and 95, respectively. At the end of the deposition process, the system was equilibrated at 300~K for 50~ps.

The second stage was annealing the deposited Ni-doped $\mathrm{MoS_2}$ film, following a simulation process similar to that used previously for the crystallization of un-doped $\mathrm{MoS_2}$.~\cite{ponomarev2022ReaxFF-MoS2-new}
At this stage, the constraints on the lowermost atoms were removed and the simulation cell height was set to the height of the reflective wall at 4.8~nm with the boundary condition in the $z$-direction changed to periodic to model bulk material.
The annealing process was carried out by heating the model at the end of the deposition stage to 5000~K over 50~ps at a ramping rate of 100~K/ps.
The structure was equilibrated at high temperature for 50~ps.
Then, the structure was cooled to 2000~K at a rate of 30~K/ps followed by equilibration for 300~ps to trigger nucleation.
Finally, the model was cooled to 300~K over 170~ps (10~K/ps). 

\begin{figure}[H]
\centering
{\includegraphics[width=0.95\textwidth]{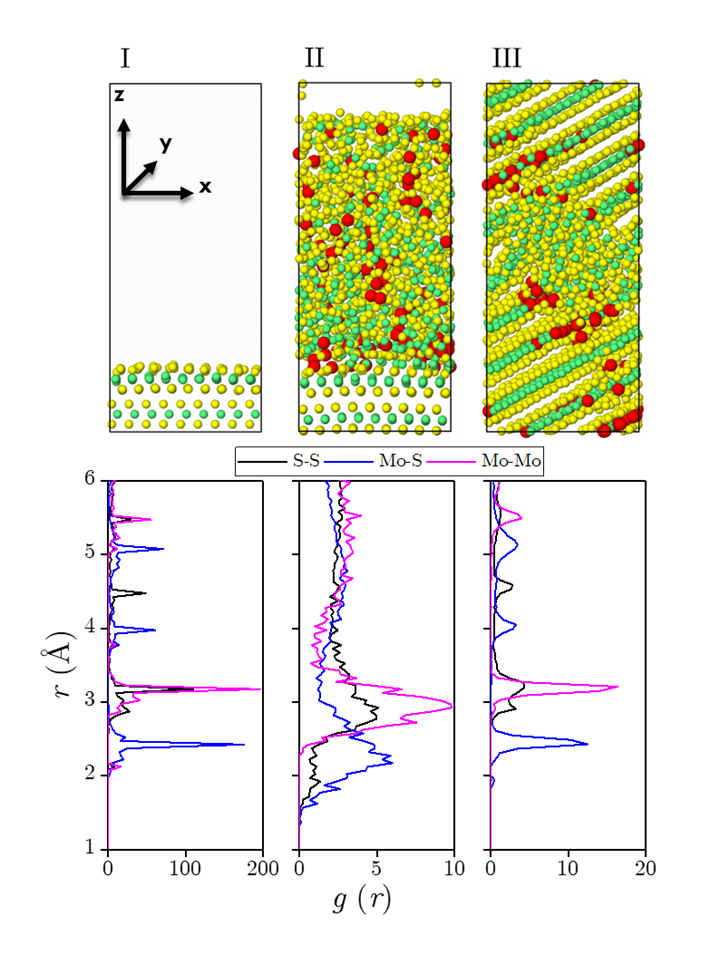}}\\
\caption{Side-view snapshots (top) and radial distribution function (bottom) of the model system,  (I) before deposition, (II) at the end of the deposition, and (III) at the end of the annealing stage. A clear transition from amorphous to crystalline is observed (II) $\rightarrow$ (III). Spheres represent S (yellow), Mo (green), and Ni (red) atoms. The black lines in the top snapshots correspond to the periodic boundaries of the cell during the annealing stage.}
\label{fig:deposition}
\end{figure}  

Snapshots of the model system before deposition, after deposition, and after annealing are shown in the top panel of Figure~\ref{fig:deposition}(I), (II), and (III), respectively. 
Visually, the model system appears amorphous at the end of deposition, consistent with experimental observations for magnetron-sputtered $\mathrm{MoS_2}$,~\cite{mcconney2016MoS2-sputter-amorphous,sirota2019MoS2-sputtering-exp} but then is mostly crystalline after annealing, as observed after annealing in experiments.~\cite{mcconney2016MoS2-sputter-amorphous,sirota2019MoS2-sputtering-exp}
Note that the middle region of the material (Figure~\ref{fig:deposition}(III)) does not appear crystalline from this view, but is in fact crystalline at an angle relative to the perspective shown here.
An alternate angle from which the crystallinity of the entire model is visible is shown in Fig. S16.
The crystallization process can be quantified using radial distribution functions (RDFs) of S--S, S--Mo, and Mo--Mo atom distances at the end of each stage of the simulation. 
The RDF of the initial crystal substrate after equilibration, shown in Figure~\ref{fig:deposition}(bottom panel (I)), exhibits clear peaks indicative of a perfect crystal.
At the end of the deposition, Figure~\ref{fig:deposition}(bottom panel (II)) shows broad close-distance peaks and only weak further-distance (second neighbor) peaks, indicating an amorphous structure.
Then, the RDF after annealing, shown in Figure~\ref{fig:deposition}(bottom panel (III)), has the regular peaks again, only slightly broadened from the before-deposition peaks, confirming that the material is in fact crystalline.

\begin{figure}[H]
\centering
{\includegraphics[width=0.95\textwidth]{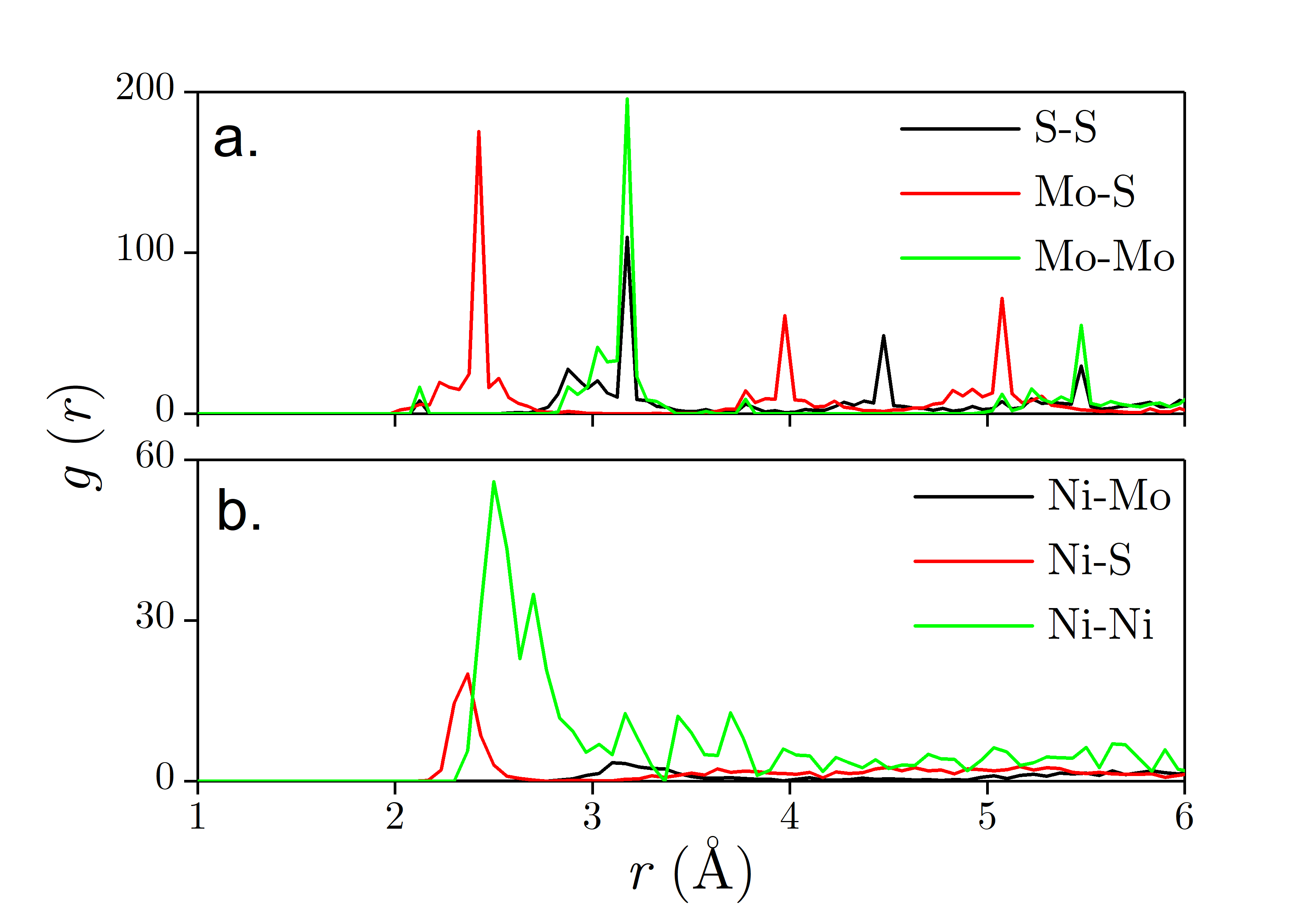}}\\
\caption{Radial distribution functions for (a) 2H $\mathrm{MoS_2}$ substrate before deposition and (b) system after annealing. Similar peak locations for Ni--S and Mo--S, and for Ni--Mo and Mo--Mo, indicate occupation of Mo sites by Ni after annealing.}
\label{fig:RDF-annealed-and-2H}
\end{figure}

Visual analysis of the simulation after annealing suggested that most of the Ni atoms positioned themselves at Mo sites, resulting in a Mo-substituted Ni-doped $\mathrm{MoS_2}$ structure.
To confirm this, the distribution of Ni--S and Ni--Mo distances at the end of annealing are shown in Figure~\ref{fig:RDF-annealed-and-2H}(b).
The heights of the first Ni--S (red) peaks after annealing are at similar distances to those observed in Figure~\ref{fig:RDF-annealed-and-2H}(a) for the Mo--S (red) peak. We can conclude that Ni atoms are bonded to S more than to Mo, consistent with the Mo-substituted Ni-doped $\mathrm{MoS_2}$ structure. 
Previous DFT calculations have shown that, under S-rich conditions, Mo-substituted is the most favorable doping location for Ni.~\cite{guerrero2020phase}
Visual analysis of the post-annealing simulations also indicated that the Ni atoms were not randomly distributed throughout the crystal, but rather formed few-atom clusters. This observation is consistent with phase separation predicted for Mo substitution according to convex hull analysis of DFT calculations.\cite{guerrero2020phase}
Similar behavior has been previously observed for gold atoms co-sputtered with $\mathrm{MoS_2}$,~\cite{Scharf2013-nx} which is further support for the physical realism of these simulations and the new force field.

\section{Conclusion}
Two new ReaxFF force fields were developed for Ni-doped $\mathrm{MoS_2}$. 
The force fields were developed by adding the parameters for Ni-Mo-S, Ni-S-Mo, and S-Ni-Mo angles to a previously optimized force field for $\mathrm{MoS_2}$,~\cite{ponomarev2022ReaxFF-MoS2-new, Ostadhossein2017} and tuning those parameters to match DFT-calculated energies.
The parameterization was based on strained DFT calculations of Mo-substituted, S-substituted, octahedral, and tetrahedral intercalation structures of the Ni dopant.
Both force fields showed accurate and reliable results with the force field that was based on the 2022 parameters performing slightly better.
Hence, the results from the 2022 parameters were shown in the main text.
However, since the two original parameter sets were optimized for different $\mathrm{MoS_2}$ models, i.e., single-layer versus bulk, the 2017 parameters' equation of state results were included in Figs. S5-S9.
The final force field developed based on the 2022 potential was able to accurately predict the energy difference between tetrahedral and octahedral intercalation, lengths of Mo--Ni and S--Ni bonds, lattice constants, S--S distance, and interlayer separation.
Furthermore, the developed force field agreed with DFT on the relaxed geometries of Ni-doped $\mathrm{MoS_2}$ structures with vacancies. 
We note that the force field was not trained for interlayer sliding,\cite{Guerrero_sliding} and our initial testing indicates that it was not able to accurately capture sliding behavior for most dopant configurations; improvement of the force field to capture sliding energies could be considered in future work.
To perform simulations relevant to catalysis with Ni-doped $\mathrm{MoS_2}$, a next step would be to incorporate interactions with other elements like H. In addition, including torsion terms and training of the bond angle parameters that were taken from the literature might improve the accuracy of the force field.
However, the force field is robust for modeling the crystal structures of Ni-doped $\mathrm{MoS_2}$ and their elastic behavior, as well as the phase transition between amorphous and crystalline, and also the underlying mechanisms of doping.
The ReaxFF force fields developed in this work will enable future simulation-based studies of the fundamental mechanisms by which Ni dopants affect $\mathrm{MoS_2}$. 

\begin{suppinfo}
Comparisons of the Ni/S and Mo/Ni force fields with DFT training data, comparison between ReaxFF and DFT equations of state with the force field developed based on the 2017 ReaxFF $\mathrm{MoS_2}$ parameters~\cite{Ostadhossein2017}, sliding potentials for doped and pristine $\mathrm{MoS_2}$ with 2017 and 2022 parameters, and alternate snapshot of the post-annealing doped material (PDF); relaxed structures from DFT (.POSCAR) and ReaxFF (.XYZ) and their minimized energies; and ReaxFF parameter files (.txt) developed based on the 2017~\cite{Ostadhossein2017} and 2022~\cite{ponomarev2022ReaxFF-MoS2-new} $\mathrm{MoS_2}$ parameters.
\end{suppinfo}

\section{Acknowledgments}
We acknowledge Rijan Karkee for providing Ni-doped 1T and 1H structures. Computing resources were provided by the Multi-Environment Computer for Exploration and Discovery (MERCED) cluster at UC Merced, funded by National Science Foundation Grant No. ACI-1429783, and the National Energy Research Scientific Computing Center (NERSC), a U.S. Department of Energy Office of Science User Facility operated under Contract No. DE-AC02-05CH11231. This work was supported by the Merced nAnomaterials Center for Energy and Sensing (MACES), a NASA-funded research and education center, under awards NNX15AQ01 and NNH18ZHA008CMIROG6R, and by UC Merced start-up funds.
\bibliography{references}
\pagebreak

\section*{TOC Graphic}
\label{For Table of Contents Only}
\centering
\includegraphics[]{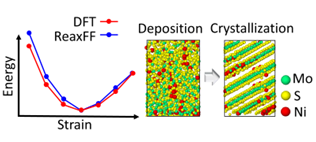}

\end{document}